\documentclass[10pt,journal,compsoc]{IEEEtran}
\ifCLASSOPTIONcompsoc
  \usepackage[nocompress]{cite}
\else
  \usepackage{cite}
\fi
\ifCLASSINFOpdf
\else
\fi
\usepackage[cmex10]{amsmath}
\usepackage{amssymb}
\usepackage{array}
\usepackage[pdftex]{graphicx}
\usepackage{multirow}
\usepackage{xcolor}
\usepackage[tight,footnotesize]{subfigure}
\usepackage{algorithm} 
\usepackage{algorithmic} 
 \usepackage[cmex10]{amsmath}
\usepackage{amssymb}
\usepackage{booktabs}
\usepackage{graphicx}
\usepackage{url}
\newtheorem{theorem}{Theorem}

\newcolumntype{C}[1]{>{\vspace{0.5em}\begin{minipage}{#1}\centering\let\newline\\
\arraybackslash\hspace{0pt}}m{#1}<{\end{minipage}\vspace{0.5em}}}

\begin{document}
%
\title{Optimal Response to Burstable Billing \\ under Demand Uncertainty}
%
%
%
%

\author{Yong~Zhan,~\IEEEmembership{Student Member,~IEEE,}
        Mahdi~Ghamkhari,~\IEEEmembership{Student Member,~IEEE,}
        Hossein~Akhavan-Hejazi,~\IEEEmembership{Student Member,~IEEE,}
        Du~Xu,~\IEEEmembership{Member,~IEEE,}
        and~Hamed~Mohsenian-Rad,~\IEEEmembership{Senior Member,~IEEE}
\IEEEcompsocitemizethanks{\IEEEcompsocthanksitem Y. Zhan and D. Xu are with the Key Laboratory of Optical Fiber Sensing and Communications, University of Electronic Science and Technology of China, Chengdu, China. \protect\\
E-mail: \{yzhan.china, xudu.uestc\}@gmail.com.
\IEEEcompsocthanksitem M. Ghamkhari, H. Akhavan-Hejazi and H. Mohsenian-Rad are with the Department of Electrical Engineering, University of California, Riverside, CA, USA. \protect\\
E-mail: \{ghamkhari, shejazi, hamed\}@ece.ucr.edu.}
\thanks{This work was done when Y. Zhan was a Visiting Student at the Smart Grid Research Lab, University of California at Riverside. The corresponding author is H. Mohsenian-Rad.}}

\maketitle
\begin{abstract}
Burstable billing is widely adopted in practice, e.g., by colocation data center providers, to charge for their users, e.g. data centers, for  transferring data. However, there is still a lack of research on what the best way is for a user to manage its workload in response to burstable billing. To overcome this shortcoming, we propose a novel method to optimally respond to burstable billing under demand uncertainty. First, we develop a tractable mathematical expression to calculate the \emph{95th percentile usage} of a user, who is charged by a provider via burstable billing for bandwidth usage. This model is then used to formulate a new bandwidth allocation problem to maximize the user's surplus, i.e., its net utility minus cost.
Additionally, we  examine different non-convex solution methods for the formulated stochastic optimization problem.
 We also extend our design to the case where a user can receive service from multiple providers, who all employ burstable billing. 
 Using real-world workload traces, we show that our proposed method can reduce user's bandwidth cost by $26\%$ and increase its total surplus by $23\%$,  compared to the current practice of allocating bandwidth on-demand.
\end{abstract}




%
\IEEEpeerreviewmaketitle

\section{Introduction}\label{sec:introduction}


%
%
%
%
\IEEEPARstart{B}{urstable} billing, is a smart data pricing (SDP) method that is used in practice, e.g., by Internet service providers, to charge for transferring data \cite{Odlyzko2001,Dimitropoulos2009,ReddyvariRaja2014,Sathiaseelan2015}. Recently, burstable billing is also widely adopted by Colocation Data Center (CDC) providers, e.g., Creative Data Concepts \cite{cdc:bandwidth}, NetSource Communications \cite{netsource:bandwidth} and Co-Location.com \cite{co-location.com:bandwidth},  as a means to charge their users for bandwidth usage.
According to Colocation America, bandwidth billing has become the second largest aspect of CDC users' overall costs, second to energy billing \cite{colocationamerica:bandwidth}.

Under burstable billing, the \emph{provider}, who provides its \emph{users} with links for data transferring, will measure each of its user's usage of bandwidth based on the user's \emph{peak usage} at a certain percentile, often at the \emph{95th percentile usage}. By construction, burstable billing neglects the user's usage of bandwidth during any time other than period of peak use. Hence, burstable billing allows users to exceed their usage thresholds for a short period without facing financial penalty \cite{ReddyvariRaja2014}.

In general, burstable billing can be studied from two different viewpoints: \emph{provider's} and \emph{user's}. For studies that address burstable billing from the provider's viewpoint \cite{Dimitropoulos2009,ReddyvariRaja2014,Raja2014,Clegg2014}, a common strategy is for the provider to move different users' workloads across space and time to avoid coinciding their peak usages, thus, reducing the overall peak demand for bandwidth \cite{Clegg2014}. However, whether or not users are willing to modulate their workloads is often overlooked.

\begin{figure}[t]
	\centering
	\includegraphics[width=0.9\linewidth]{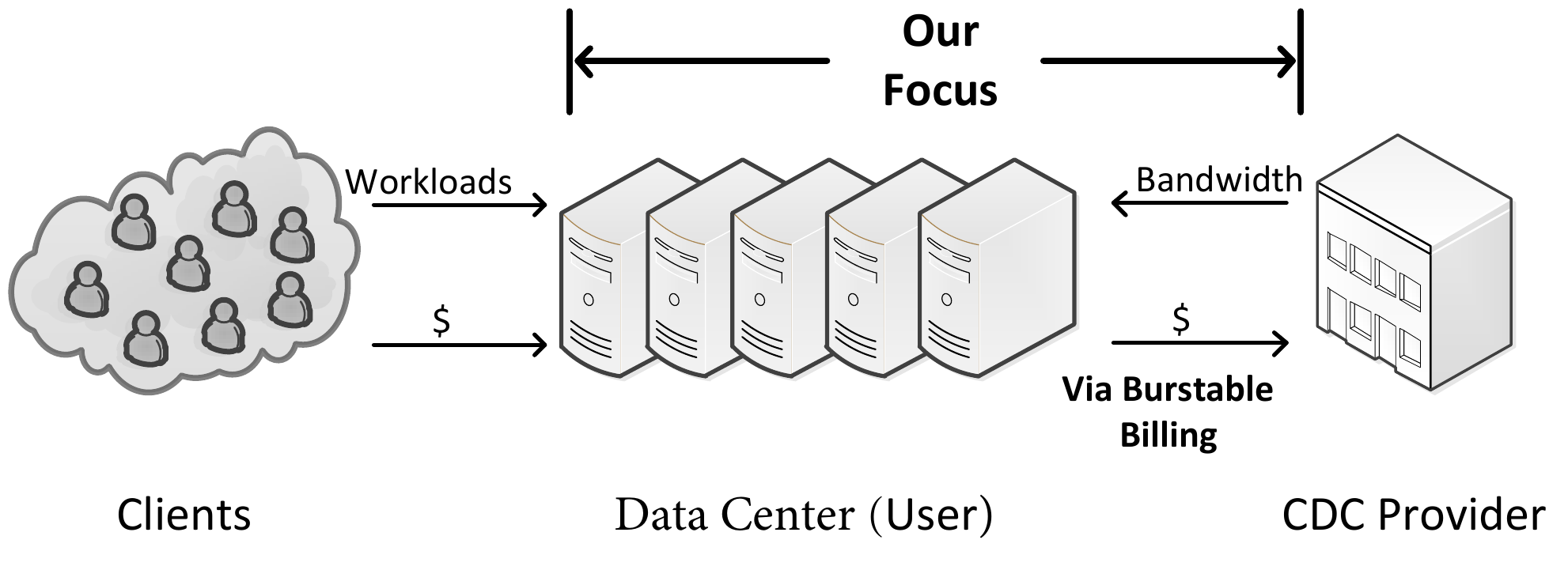}
	\caption{An example setup for the application of burstable billing: a data center, i.e., user of a CDC provider, utilizes bandwidth provided by the CDC provider to serve outside clients with uncertain demands.} \vspace{-0.2cm}\label{fig:example}
\end{figure}


The studies that address burstable billing from the user's perspective have emerged only recently. So far, due to the lack of a tractable mathematical expression to calculate the \emph{95th percentile usage} of bandwidth, a common approach has been to use experimental and/or heuristic methods, e.g., as in \cite{Traverso2015,Golubchik2013,Laoutaris2011, Nandagopal2012, Marcon2011, Stanojevic2011}. There are also few studies that are analytical; however, they assume that the workload has a specified distribution, e.g., Gaussian distribution \cite{Xu2012}, or they focus on peak pricing, i.e., the $100$ percentile billing instead of $95$ percentile billing \cite{Zhang2014}, or they assume that the cost of bandwidth is volume-based \cite{Xu2013,Xiang2014a}.


In this paper, as illustrated by Fig. \ref{fig:example},  we are interested in studying burstable billing from the \emph{user's viewpoint} by taking into consideration the trade-off between cost and performance based on user's preferences. Specifically, we seek to answer this fundamental question: \emph{What is the best way for an individual user, such as a data center in a CDC, who is charged via burstable billing, to manage its operation and the use of bandwidth}? Our approach to answer this question is based on formulating and solving an optimization problem for bandwidth usage which aims at maximizing the user's \emph{surplus}, i.e., its net utility minus cost.
%
%

We take into consideration the fact that, in practice, neither the user nor the provider have perfect knowledge about the workload, and thus the demand for bandwidth in the future. For example, when it comes to a user in a CDC as in Fig. \ref{fig:example}, the workload is initiated by the user's clients, not the user itself. Therefore, in our analysis, we address demand uncertainty within a stochastic optimization framework.


The main contributions of this paper are as follows:

\vspace{0.05cm}

\begin{enumerate}
  \item To the best of our knowledge, this is the first paper to study the problem of optimal responding to burstable billing from a single user's viewpoint under demand uncertainty with arbitrary probability distributions. 


\vspace{0.05cm}

  \item To facilitate the use of systematic optimization, we develop a tractable mathematical expression to calculate the \emph{95th percentile usage} of bandwidth. This model is then used to formulate a novel bandwidth allocation problem to maximize the user's surplus.
Additionally, we  examine different solution methods to find the exact and near-optimal solutions of the formulated problem.


\vspace{0.05cm}

  \item We extend our design  as well to  another emerging practical scenario where a user can receive service from multiple  providers, e.g., when a user can request content it needs from multiple providers that all employ burstable billing. Accordingly, our problem formulation also addresses workload distribution in addition to bandwidth allocation.


\vspace{0.05cm}

  \item We evaluate our design based on a real-world workload trace: Wikipedia Page View data \cite{Wikipageview}. With a typical  workload forecasting method, we show that the use of our design is particularly rewarding if a user is charged by high bandwidth price and/or it is more sensitive to price than to performance. Finally, we also show the advantage of utilizing services from multiple providers, where we can further increase the user's surplus by distributing its workload to multiple providers that employ burstable billing.
\end{enumerate}

 \begin{figure}[!t]
	\centering
	\includegraphics[width=0.8\linewidth]{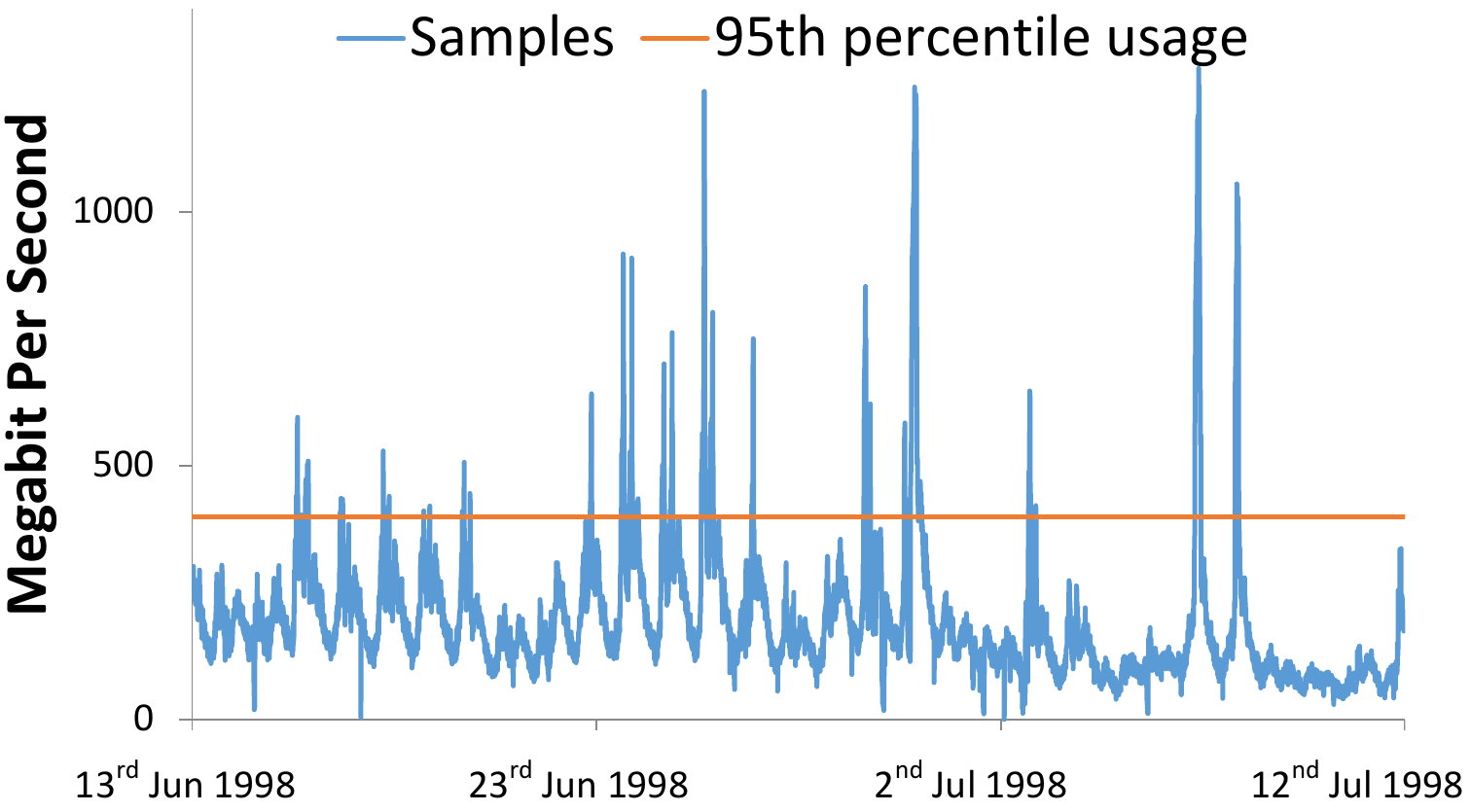}
	\caption{An example for calculating the \emph{95th percentile usage}:
a total of 8640 samples are collected for a user during one billing cycle. After throwing away the top
$5 \%$, i.e.,  $5* 8640 / 100 = 432$ samples, the \emph{95th percentile usage}
is obtained as 399.1277 Mbps, which is equal to the highest recorded bandwidth usage  of the remaining $95 *8640 / 100 = 8208$ samples.
The \emph{95th percentile usage} is shown by the red line.
Here, the user is allowed to have a
total of 432 bursts above the red line without facing financial penalty.}\label{fig:95th}
\end{figure}
\section{Problem Formulation}
In this section, we formulate a mathematical expression for a  user's \emph{95th percentile usage}, which is a key concept in burstable billing. This model is then used to obtain the user's expected bandwidth cost and surplus prior to a billing cycle.

\subsection{95th Percentile Usage}\label{subsec:95thpercentileusage}
In order to apply burstable billing, a provider first divides a
billing cycle into $\tau$  time intervals of equal length $T$. The
length of time intervals could be as low as 30 seconds,
though typically the time intervals of $T = 5$ minuets are
considered \cite{Dimitropoulos2009}. Next, to obtain a user's \emph{95th percentile usage}, the provider takes
 samples of the user's usage of bandwidth, e.g., once every five minutes during  that billing cycle. Then, the top $5\%$ of the samples gathered within the billing cycle are thrown away and  the highest element of the remaining $95\%$ samples is taken as the user's \emph{95th percentile usage}. An example for calculating the \emph{95th percentile usage} is shown in Fig. \ref{fig:95th}. Similarly, the user can obtain its own \emph{95th percentile usage}, denoted by $\mu_{95}(x[t])$, given the usage samples $x[1], \cdots, x[\tau]$ from the mathematical expression provided in the following theorem:  

\vspace{0.1cm}

\begin{theorem} \label{theorem1}
Given $x[t]$, $\forall t=1,\ldots, \tau$  as the $\tau$ samples of the bandwidth usage for a user during a billing cycle, we can model the \emph{95th percentile usage} for that user as
\begin{equation}
\label{mu_95_Model}
\begin{aligned}
& \mu_{95}(x[t]) = \!\!\! & \min_\rho \ & \max_t  \rho[t]  x[t]\\
& & \text{s.t.} \ \ & \rho[t] \in \{0,1\}, \quad \forall t, \\
& & & \! \sum_{t=1}^{\tau} {\rho[t]}=\lceil0.95 \tau\rceil,\\
\end{aligned}
\end{equation}
where the variables in the above minimization are $\rho[t]$ for all $t = 1,\ldots, \tau$, and $\lceil \cdot \rceil$ denotes the ceiling function. %
\end{theorem}

\emph{Proof:} Let us define $\widehat{\rho}[1], \ldots, \widehat{\rho}[\tau]$ such that $\widehat{\rho}[t]=0$ for each time slot $t$ at which $x[t]$ is within the top 5\% of the values in array $x[1], \ldots,x[\tau]$, and $\widehat{\rho}[t]=1$ otherwise. Clearly, we have
\begin{equation}
\mu_{95}(x[t])=\max_t \{x[t]\widehat{\rho}[t]\},
\end{equation}
\noindent which in this case, Theorem \ref{theorem1} holds.
Next, we  note that $\widehat{\rho}[1], \ldots, \widehat{\rho}[\tau]$ is a feasible solution to problem (\ref{mu_95_Model}). To complete the proof, we  show that $\widehat{\rho}[1], \ldots, \widehat{\rho}[\tau]$ is in fact the optimal solution of the minimization problem in (\ref{mu_95_Model}). We prove this by contradiction. Suppose $\tilde{\rho}[1], \ldots, \tilde{\rho}[\tau]$ is the optimal solution of problem (\ref{mu_95_Model}), where for at least one time slot $t$, we have $\tilde{\rho}[t]\neq \widehat{\rho}[t]$. Due to the equality constraint in (\ref{mu_95_Model}), 95\% of the variables in $\tilde{\rho}[1], \ldots, \tilde{\rho}[\tau]$ are equal to one. Therefore, $\tilde{\rho}[1], \ldots, \tilde{\rho}[\tau]$ could be different from $\widehat{\rho}[1], \ldots, \widehat{\rho}[\tau]$ only if there exists a time slot $t$ for which $\tilde{\rho}[t] = 1$ even though $x[t]$ \emph{is} within the top 5\% of the values in array $x[1], \ldots,x[\tau]$. In that case, we must have
%
%
%
%
%
\begin{equation}\label{formula:Core}
\max_t \{x[t]\tilde{\rho}[t]\} \geq \max_t \{x[t]\widehat{\rho}[t]\}.
\end{equation}
%
Also, since $\tilde{\rho}[1], \ldots, \tilde{\rho}[\tau]$ is assumed to be the optimal solution of problem (\ref{mu_95_Model}), by definition of optimality, we must have
%
\begin{equation}\label{formula:Core1}
\max_t \{x[t]\tilde{\rho}[t]\} \leq \max_t \{x[t]\widehat{\rho}[t]\}.
\end{equation}
From (\ref{formula:Core}) and (\ref{formula:Core1}), we can conclude that
\begin{equation}
\max_t \{x[t]\tilde{\rho}[t]\} = \max_t \{x[t]\widehat{\rho}[t]\}.
\end{equation}
However, this contradicts the assumption that $\widehat{\rho}[1], \ldots, \widehat{\rho}[\tau]$ is not optimal. Therefore, $\widehat{\rho}[1], \ldots, \widehat{\rho}[\tau]$ is the optimal solution.  $\hfill \blacksquare$

\vspace{0.2cm}


  $\rho[t]$ in problem (\ref{mu_95_Model}) is an auxiliary  variable. For each time slot $t$, if $\rho[t]=0$, it indicates that its corresponding usage, $x[t]$, is within the top $5\%$ of the values in $x[1],\ldots,x[\tau]$, thus, $x[t]$ has no impact on the \emph{95th percentile usage} $\mu_{95}(x[t])$, i.e.,  the user can  utilize bandwidth on-demand without extra cost. On the contrary, if $\rho[t]=1$, the user may restrict its usage at this time slot to reduce its \emph{95th percentile usage}.

\subsection{Cost of Bandwidth under Burstable Billing}\label{subsec:problemformulation}

Next, we formulate the user's bandwidth cost given the bandwidth usage samples $x[1],\ldots,x[\tau]$ based on burstable billing as
\begin{equation}
\label{C_95_Model}
C_{95}(x[t])= \delta \cdot \mu_{95}(x[t]),
\end{equation}
where $\delta$ (\$/Mbps) denotes the price of bandwidth under burstable billing. 
Note, the price of bandwidth $\delta$ can vary with the length of billing cycle. However, in this paper, we assume that the length of each billing cycle is fixed, i.e., the price of bandwidth $\delta$ is constant.

\subsection{Expected Surplus Prior to a Billing Cycle}\label{subsec:netutility}

Consider a user that aims to plan for its bandwidth usage \emph{prior} to a billing cycle. A key question is how to model the \emph{expected surplus}, i.e., net utility minus cost,  under uncertain demand.
Therefore, in this section we formulate  the user's expected net utility and surplus prior to a billing cycle. 

Let $D[t]$ (Mbps) be the user's demand for bandwidth at time slot $t$, which is the amount of bandwidth user needs  to fully satisfy its clients, i.e.,   to obtain the highest net utility. Note that, the user may not know its exact   demand in the future, rather  has a distribution for its demand, i.e.,   $D[t]$ is a  random variable.
Next, we note that  the user may not always choose to serve its full demand for bandwidth at a given time. 
Let $X[t]$ (Mbps) be the \emph{planned usage} of bandwidth  during time interval $t = 1, \ldots, \tau$ for the user prior to the billing cycle.
 Here, the \emph{planned usage} of bandwidth $X[t]$ is decided based on the demand $D[t]$.

We assume a general net utility function in this paper that depends only on user's bandwidth usage. At each time slot t, the utility function $U(\cdot)$ is a concave and non-decreasing function of the total bandwidth, as in \cite{Niu2012,He2012}.  However, 
the user cannot gain any extra utility by using more bandwidth than its demand. Therefore, for a billing cycle, we formulate the user's   \emph{expected net utility } as 
\begin{equation}
\label{equ:net utility}
R =\sum_{t=1}^{\tau} {\mathbb{E} (U(T\min\{X[t],D[t]\}))}
\end{equation}
corresponding to its planned usage samples $X[1], \ldots,X[\tau]$.
Here, $\mathbb{E} (.)$ denotes mathematical expectation.

 From the optimization-based model in (\ref{mu_95_Model}), one can calculate the   \emph{95th percentile usage} for each billing cycle, which is denoted by $\mu_{95}(X[t])$, as a function of \emph{planned usage} samples $X[1], \ldots,X[\tau]$. Further, the corresponding   bandwidth cost at each billing cycle can be calculated via (\ref{C_95_Model}).

From (\ref{C_95_Model}) and (\ref{equ:net utility}), the user's \emph{expected  surplus}, i.e., its  expected net utility minus its bandwidth cost during a billing cycle, is obtained as
\begin{equation}\label{equ:surplus}
S=\sum_{t=1}^{\tau} {\mathbb{E} \left(U(T\min\{X[t],D[t]\})\right)}-\delta \cdot \mu_{95}(X[t]).
\end{equation}

\section{Surplus Maximization}\label{subsec:surplusmaximization}
In this section, we aim to optimally plan the user's bandwidth usage  prior to a billing cycle to achieve the highest  	surplus.
In other words, we formulate the problem to obtain the optimal \emph{planned usage} so as to maximize the \emph{expected surplus}.
Typically, neither the user nor the provider have perfect knowledge about the user's demand for bandwidth in an upcoming billing cycle, i.e.,  $D[1], \ldots, D[\tau]$ are often uncertain. 
Here, we assume that the predictions of user's demand $D[t]$ are given, which could be either deterministic values or stochastic probability functions.  
Accordingly, we formulate the optimization problems  of maximizing the user's surplus \emph{prior} to a billing cycle under deterministic and stochastic prediction of $D[t]$.


\subsection{Surplus Maximization with Deterministic Prediction}\label{subsec:deter}
If the prediction of demand for bandwidth is deterministic, i.e., parameters $D[1], \ldots,$ $D[\tau]$ are \emph{deterministic}, from (\ref{mu_95_Model}) and (\ref{equ:surplus}), we formulate the optimization problem to maximize the user's  surplus over a billing cycle as:
\begin{equation}\label{equ:problem}
\begin{aligned}
& \underset{X[t],\rho[t]}{\text{max}} \!\!\!\!\!\!\
& &  \sum_{t=1}^{\tau} {U(T\min\{X[t],D[t]\})}-\delta \max_{t} \rho[t]X[t]\\
& \text{s.t.} & & X[t]\geq 0, \quad \quad \quad \quad \quad \forall t, \\
&&& \rho[t] \in \{0,1\}, \quad\quad\quad\ \ \:\! \forall t, \\
&&& \sum_{t=1}^{\tau} {\rho[t]}=\lceil0.95\tau\rceil.
\end{aligned}
\end{equation}
Here,
$X[t]$ is the principal variable while $\rho[t]$ is the auxiliary variable that is used to calculate the \emph{expected} \emph{95th percentile usage} as explained in Theorem \ref{theorem1}. Note that, since the net utility function does not depend on the auxiliary variable $\rho[t]$, and also because price parameter $\delta$ is nonnegative, if the principal variable $X[t]$ is set to be fixed, then the maximization in (\ref{equ:problem}) over $X[t]$ and $\rho[t]$ reduces to the minimization in (\ref{mu_95_Model}) over $\rho[t]$. Therefore, it is guaranteed that once we solve the problem in (\ref{equ:problem}), the choice of auxiliary variable $\rho[t]$ is automatically selected in a way that $\mu_{95}(X[t])$ is calculated as in (\ref{mu_95_Model}).



\subsection{Surplus Maximization with Stochastic Prediction}\label{subsec:sto}
%
Another common approach in addressing uncertainty is to obtain a probability mass function \cite{probabilitymassfunction} for each random parameter using historical workload data. This can be done in various levels of details and accuracy, e.g., see \cite{Jedynak2005}. 
In such case, we assume that each $D[t]$ shall be expressed by  $K_{t}$ possible realizations: $D_{1}[t],\ldots,D_{K_{t}}[t]$, where each realization $D_{k}[t]$ may occur with probability $\pi_{k,t}$. We have
%
%
\begin{equation}\label{equ:simplefunction}
\sum_{k=1}^{K_{t}} \pi_{k,t}=1, \quad \forall t.
\end{equation}
%
Once we use the above modeling method, we can then formulate the \emph{stochastic optimization} problem to maximize the user's expected surplus over a billing cycle as:
\begin{equation}\label{equ:problemsto}
\begin{aligned}
& \underset{X[t],\rho[t]}{\text{max}} \!\!\!\!\!\!\
& &  \sum_{t=1}^{\tau} {\sum_{k=1}^{K_{t}} {\pi_{k,t} U(T\min\{X[t],D_k[t]\})}}-\delta \max_{t}\rho[t]X[t]\\
& \text{s.t.} & & X[t]\geq 0, \quad \quad \quad \quad \quad \forall t, \\
&&& \rho[t] \in \{0,1\}, \quad\quad\quad\ \ \:\! \forall t, \\
&&& \sum_{t=1}^{\tau} {\rho[t]}=\lceil0.95\tau\rceil.
\end{aligned}
\end{equation}

\vspace{-0.1cm}

\section{Solution Method}

Both problems (\ref{equ:problem}) and (\ref{equ:problemsto}) are nonlinear, mixed-integer programmings, which are generally considered to be hard problems to solve. Nevertheless, in this section, we explain how these problems can be solved with reasonable computational complexities.



\subsection{Deterministic Problem}\label{subsec:deterministicproblem}
For the deterministic problem (\ref{equ:problem}),
we can intuitively obtain the optimal solution for  variables $\rho[1], \ldots, \rho[\tau]$ without numerically solving the problem. This property can be expressed mathematically in the following theorem.


\vspace{0.1cm}
\begin{theorem} \label{theorem2}
Let $\vartheta$ denote the set of all time slots $t$ at which $D[t]$ is within the top $5\%$ of the values in $D[1],\ldots,D[\tau]$.

\vspace{0.2cm}

(a) There exists an optimal solution for  the deterministic problem (\ref{equ:problem}) in which the values of auxiliary variables $\rho[1], \ldots, \rho[\tau]$  are as follows:
\begin{equation}\label{equ:rhooptimal}
\rho^{\star}[t]=
\begin{cases}
0, &\forall  t \in \vartheta;\\
1, & \text{otherwise}.
\end{cases}
\end{equation}
(b) Once  the optimal values of $\rho$  in the deterministic problem (\ref{equ:problem}) are replaced from  (\ref{equ:rhooptimal}), the solution for the principal variables $X[1], \ldots, X[\tau]$ of  the deterministic problem (\ref{equ:problem}) are obtained from the following convex optimization problem:
\begin{equation}\label{equ:deter}
\begin{aligned}
& \underset{X[t]}{\text{max}} & &\sum_{t=1}^{\tau} U(TX[t])-\delta \max_t \rho^{\star}[t] X[t] \\
& \text{s.t.} & &0 \leq X[t] \leq D[t],   \ \forall  t,
\end{aligned}
\end{equation}
where $\rho^{\star}[t]$ is given by (\ref{equ:rhooptimal}).
\end{theorem}

\emph{Proof:} First, one can easily find that the objective function in the deterministic problem (\ref{equ:problem}) is a non-increasing function of $X[t]$, when $X[t]\geq D[t]$. Therefore, the optimization problem (\ref{equ:problem}) can be reformulated  as 
\begin{equation}\label{equ:problem11}
\begin{aligned}
& \underset{X[t],\rho[t]}{\text{max}} \!\!\!\!\!\!\
& &  \sum_{t=1}^{\tau} {U(TX[t])}-\delta \max_{t}\rho[t]X[t]\\
& \text{s.t.} & & 0 \leq X[t] \leq D[t], & \forall t, \\
&&& \rho[t] \in \{0,1\},& \forall t, \\
&&& \sum_{t=1}^{\tau} {\rho[t]}=\lceil0.95\tau\rceil.
\end{aligned}
\end{equation}
Next, we note that $\rho^\star[t]$ in (\ref{equ:rhooptimal}) is a feasible solution for the problem (\ref{equ:problem11}). Let $\bar{\vartheta}$ denote the complement set of $\vartheta$, i.e., $\bar{\vartheta}=\{1,\ldots,\tau\}-\vartheta$. Let $\rho^{c}[t]$ denote the true optimal solution of $\rho[t]$ for the  problem (\ref{equ:problem11}). The  solution of  usage  $X^{\star}[t]$, obtained from  (\ref{equ:problem11}) by setting $\rho[t]=\rho^\star[t]$,  is as follows:
\begin{equation}\label{equ:lemmasatisfied}
X^{\star}[t]=
\begin{cases}
D[t], &\forall t \in \vartheta;\\
\min\{\mu^{\star},D[t]\}, &\forall t \in \bar{\vartheta},\\
\end{cases}
\end{equation}
where $\mu^{\star}$ is the optimal \emph{95th percentile usage} of bandwidth corresponding to $X^{\star}[t]$.

To complete the proof of theorem we only need to show that $\rho^c[t]=\rho^\star[t]$. 
Next, We prove by contradiction that this argument indeed holds.  In other words, if   we assume that $\rho^{c}[t] \neq \rho^\star[t]$, then the user's total surplus with $\rho^{c}[t]$ will be less than the user's surplus with $\rho^\star[t]$. 

Let  $\rho^{c}[t] \neq \rho^\star[t]$ so that:

\begin{equation}\label{formula:rhoc}
\rho^c[t]=
\begin{cases}
0,  &\forall t \in \nu; \\
1. &\forall t \in \bar{\nu},\\
\end{cases}
\end{equation}
where $\nu$ is some set so that $\nu\neq\vartheta$ and  $\bar{\nu}$ is the complement set  of $\nu$. This assumption implies that, for at least one time slot $t \in \nu$, $D[t]$ is \emph{not} within the top $5\%$ of the values in array $D[1],\ldots,D[\tau]$. The optimal  usage of bandwidth $X^{c}[t]$ in this case becomes
%
\begin{equation}\label{equ:scenario1}
X^{c}[t]=
\begin{cases}
D[t], &\forall t \in  \nu \\
\min\{\mu^{c},D[t]\}, &\forall t \in \bar{\nu},\\
\end{cases}
\end{equation}
where $\mu^{c}$ is the optimal \emph{95th percentile usage} of bandwidth corresponding to $X^{c}[t]$.

\noindent
We prove that $\mu^{c} \leq \max_{t \in \bar{\vartheta}}D[t]$. Considering a scenario where the user \emph{plans} to utilize bandwidth on-demand, i.e., $\forall t \in \{1,\ldots,\tau\}$, $X[t]=D[t]$. In this case, the \emph{95th percentile usage} of bandwidth becomes $\max_{t \in \bar{\vartheta}}D[t]$, which is obviously the highest feasible \emph{95th percentile usage} of bandwidth of problem (\ref{equ:problem11}). Thus, $\mu^{c} \leq \max_{t \in \bar{\vartheta}}D[t]$.

\noindent
Then, we prove that $(X^{\star}[t],\rho^{\star}[t])$ is the optimal solutions of problem (\ref{equ:problem11}). Let 
\begin{equation}\label{equ:lemmasatisfied}
X^{\star \star}[t]=
\begin{cases}
D[t], &\forall t \in \vartheta;\\
\min\{\mu^{c},D[t]\}, &\forall t \in \bar{\vartheta}.\\
\end{cases}
\end{equation}
Since $\mu^{c}\leq \max_{t \in \bar{\vartheta}}D[t]$, $\max_{t} X^{\star \star}[t] \rho^{\star}[t]=\mu^{c}$. Namely, with $(X^{\star \star}[t],\rho^{\star}[t])$, the \emph{95th percentile usage} of bandwidth equals $\mu^{c}$. In this case, we have
\begin{equation}\label{formula:Cost_Eq}
C_{95}(X^{\star \star}[t])=C_{95}(X^c[t])=\delta \cdot \mu^{c}.
\end{equation}
\noindent
Also, from (\ref{equ:net utility}), we have
\begin{equation}\label{formula:Rev1}
R(X^{\star \star}[t])=\sum_{t\in\vartheta}U(TD[t])+ \sum_{t\in\bar{\vartheta}}U(T\min\{\mu^{c},D[t]\})
\end{equation}
and
\begin{equation}\label{formula:Rev2}
R(X^c[t])= \sum_{t\in\nu}U(TD[t])+ \sum_{t\in\bar{\nu}}U(T\min\{\mu^{c},D[t]\}).
\end{equation}
%
Let $f(D[t])=U(TD[t])- U(T\min\{\mu^{c},D[t]\})$. From (\ref{formula:Rev1}) and (\ref{formula:Rev2}),  we calculate $R(X^{\star \star}[t])-R(X^c[t])$, which equals to
\begin{equation}\label{formula:Rev_Diff}
\begin{aligned}
&\sum_{t\in \vartheta-\nu} f(D[t])- \sum_{t\in \nu-\vartheta} f(D[t]).
\end{aligned}
\end{equation}

Next, note that $f(D[t])$ is in fact equal to:
\begin{equation}\label{formula:fx}
f(D[t])=
\begin{cases}
U(TD[t])- U(T\mu^{c}), & \text{if} \ D[t] \geq \mu^{c};\\
0, & \text{otherwise}.
\end{cases}
\end{equation}
\noindent
Since  $U(\cdot)$ is nondecreasing and $T \geq 0$, from (\ref{formula:fx})$, f(D[t])$ is nondecreasing, too.

\noindent
Then, we can find that
\begin{equation}\label{formula:DD}
D[t_1]\geq D[t_2]\ \ \  \forall  t_1\in \vartheta-\nu,\ \    \forall t_2\in \nu- \vartheta.
\end{equation}

\noindent
From (\ref{formula:Rev_Diff})  and (\ref{formula:DD}) and since $f(D[t])$ is nondecreasing over $D[t]$ and $\|\vartheta-\nu\|=\|\nu- \vartheta\|$, we have

\begin{equation}\label{formula:Rev_Eq}
R(X^{\star \star}[t]) \geq R(X^c[t]).
\end{equation}

\noindent
From (\ref{formula:Cost_Eq}) and (\ref{formula:Rev_Eq}), the obtained surplus with $(X^{\star \star}[t],\rho^{\star}[t])$ is no less than the one with $(X^{c}[t],\rho^{c}[t])$. Since $X^{\star }[t]$ is the optimal solution of $X[t]$ of problem (\ref{equ:problem11}) corresponding to $\rho^{\star}[t]$, the obtained surplus with $(X^{\star}[t],\rho^{\star}[t])$ is no less than the one with $(X^{\star \star}[t],\rho^{\star}[t])$. Therefore, the obtained surplus with $(X^{\star}[t],\rho^{\star}[t])$ is no less than the one with $(X^{c}[t],\rho^{c}[t])$. Since $(X^{c}[t],\rho^{c}[t])$ was assumed to be optimal,
%
$(X^{\star}[t],\rho^{\star}[t])$ is an optimal solution of problem (\ref{equ:problem11}). $\hfill \blacksquare$


\vspace{0.2cm}

From Theorem \ref{theorem2}, one can convert the non-convex problem (\ref{equ:problem}) onto a convex program (\ref{equ:deter}), which can be effectively solved using convex programming techniques \cite{Boyd2004}.

\subsection{Stochastic Problem}\label{subsec:stochasticproblem}

If parameters $D[1], \ldots, D[\tau]$ are random, then we do \emph{not} know at what time slots the burst will occur in the demand for bandwidth. Accordingly, we cannot separately figure out the optimal values of $\rho[1], \ldots, \rho[\tau]$. Therefore, we have no choice but solving the original stochastic problem (\ref{equ:problemsto}).

A key difficulty in solving the stochastic problem (\ref{equ:problemsto}) is that even if we relax the binary constraints, i.e., even if we choose $\rho[t]$ to be a continuous number between 0 and 1, the relaxed problem is still difficult to solve due to the non-convex term $\rho[t]X[t]$ in the objective function. Interestingly, we can tackle this undesirable property  as it is explained in a theorem below.

\vspace{0.1cm}
\begin{theorem} \label{theorem3}
 The stochastic problem (\ref{equ:problemsto}) is equivalent to: 
\begin{equation}\label{equ:offsto}
\begin{aligned}
& \underset{X[t],\rho[t],\phi}{\text{max}}
& &\sum_{t=1}^{\tau} {\sum_{k=1}^{K_{t}} {\pi_{k,t} U(T\min\{X[t],D_k[t]\})}}-\delta \cdot \phi\\
& \text{s.t.}
& & X[t]\leq \phi+L(1-\rho[t]), && \forall t, \\
&&& X[t] \geq 0, && \forall t, \\
&&& \rho[t] \in \{0,1\}, &&  \forall t, \\
&&& \sum_{t=1}^{\tau} {\rho[t]}=\lceil0.95\tau\rceil, \\
\end{aligned}
\end{equation}
where $L$ is a large number compared to the available bandwidth, and $\phi$ is another auxiliary variable.
\end{theorem}

\vspace{0.1cm}


\emph{Proof:} At each time slot $t$, if $\rho[t] = 0$, then the first constraint in problem (\ref{equ:offsto}) reduces to $X[t] \leq \phi + L, \ \forall t$, which always holds regardless of the values of $X[t]$ and $\phi$. If $\rho[t] = 1$, then the first constraint in (\ref{equ:offsto}) reduces to $X[t] \leq \phi, \ \forall t$. In that case, since the objective function in (\ref{equ:offsto}) is to minimize $\phi$, we necessarily obtain that  $\phi = \max_t\rho[t]X[t]$ at any optimal solution of problem (\ref{equ:offsto}). This is clearly an outcome that we intended.  $\hfill \blacksquare$

\vspace{0.2cm}

Given the equivalence of the stochastic problem (\ref{equ:problemsto}) and (\ref{equ:offsto}), we can solve problem (\ref{equ:offsto}) instead of (\ref{equ:problemsto}). 
Next, we notice that  from (\ref{equ:offsto}), once we relax the binary constraints, the relaxed problem is convex. Therefore, we can find the exact optimal solution of problem (\ref{equ:offsto}) using branch-and-bound method \cite{Lawler1966}, where at each branching step we need to solve a convex optimization problem. We refer to this approach as the convex branch-and-bound (CBB) method.

While the CBB method is effective to obtain the exact optimal solution of the stochastic  problem (\ref{equ:problemsto}), solving a nonlinear (although convex) problem at each iteration of the branch-and-bound algorithm could be time consuming. Since the nonlinearity in problem (\ref{equ:offsto}) is due to the nonlinear utility function $U(\cdot)$, one way to make problem (\ref{equ:offsto}) linear is to replace $U(\cdot)$ with its piece-wise linear approximation. This is explained in the following theorem.

\vspace{0.1cm}

\begin{theorem} \label{theorem4}
Let $N$ denote the number of tangent lines in the piece-wise linear approximation of the utility function $U(\cdot)$. If $N \rightarrow \infty$, then the problem in (\ref{equ:offsto}) is equivalent to the  mixed-integer linear optimization problem:
\begin{equation}\label{equ:offstomilp}
\begin{aligned}
& \underset{\begin{subarray}{c}X[t],\rho[t],\phi,\\ Q_{k}[t],h_{k}[t] \end{subarray}}{\text{max}}
& &\sum_{t=1}^{\tau} {\sum_{k=1}^{K_{t}} {\pi_{k,t} h_{k}[t]}}-\delta \cdot \phi\\
& \text{s.t.}
& & X[t]\leq \phi+L(1-\rho[t]), && \forall t, \\
&&& X[t] \geq 0, && \forall t, \\
&&& \rho[t] \in \{0,1\}, &&  \forall t, \\
&&& \sum_{t=1}^{\tau} {\rho[t]}=\lceil0.95\tau\rceil, \\
&&& Q_{k}[t] \leq X[t], && \forall t,k,\\
&&& Q_{k}[t] \leq D_{k}[t], && \forall t,k,\\
&&& h_k[t] \leq U(n\Delta[t])+\\
&&& \ \ \ \ \ \ \ \ \ \: U^{'}(n\Delta[t])(TQ_{k}[t]-n\Delta[t]), \quad \!\!\!\! && \forall t,k,n,
\end{aligned} \!\!\!
\end{equation}
%
%
where $n = 1,\ldots, N$. Here, $Q_{k}[t]$ and $h_k[t]$ are auxiliary variables for tangent line $k$.
\end{theorem}

\vspace{0.1cm}

\emph{Proof:} As it can be seen from (\ref{equ:offstomilp}), we first replace  $\min \{X[t],D_{k}[t]\}$ in the objective function of (\ref{equ:offsto}) with an auxiliary variable $Q_{k}[t]$. Here,   $Q_{k}[t]$ is upper bounded by $X[t]$ and $D_{k}[t]$, which is exactly the type of constraint that we need to model the min function $\min \{X[t],D_{k}[t]\}$. Next, the concave function $U(TQ_k[t])$ is replaced by a new variable $h_k[t]$. Also, as in the last constraint in (\ref{equ:offstomilp}), $h_k[t]$ is upper bounded by $N$ number of tangents lines to the concave curve $U(TQ_k[t])$. Therefore, if $N \to \infty$, $h_k[t]$ is equivalent to $U(TQ_k[t])$. Accordingly, the problem formulation in (\ref{equ:offstomilp}) becomes equivalent to the one in (\ref{equ:offsto}). $\hfill \blacksquare$

\vspace{0.2cm}

 The usefulness of problem (\ref{equ:offstomilp}) depends on the choice of parameter $N$. However, as we will see in Section \ref{subsec:impacttengentslines}, we can obtain the near exact optimal solution of the stochastic surplus maximization problem even if $N = 3$. There exist effective solvers to solve mixed-integer linear programming (MILP), such as CPLEX \cite{CPLEX}. We will see in Section \ref{subsec:impacttengentslines} that solving the MILP in (\ref{equ:offstomilp}) is computationally more tractable than the CBB method.

Before we end this section, we must point out that one can obtain an \emph{approximate} solution for problem (\ref{equ:offstomilp}) by terminating the optimization solver at certain guaranteed optimality bounds in order to significantly lower computational complexity. We will further discuss this option in Section \ref{subsec:impacttengentslines}. 

\vspace{0.1cm}

\section{Extensions and Remarks}
In this section, we discuss two interesting analysis with regards to the proposed design. 
First,  we extend our design to a scenario where a user has the option to receive service from multiple providers. An example is a user can download specified content over different transit links that is owned by different ISPs, who charge the user via burstable billing. 
Second, we show in this section that  a user can further improve its surplus by updating the usage of bandwidth in real-time, i.e. during the billing cycle, based on the newly \emph{exposed} actual demand information.

\subsection{Extension to Multiple Providers}
Let $X_{i}[t]$ denote the \emph{planned usage} of bandwidth at provider $i$ at time slot $t$ decided based on the user's demand $D[t]$. 
Let $\delta_i$ (\$/Mbps) denote the price of bandwidth at provider $i$. In this case, in each billing cycle, the  \emph{expected surplus} of the user with multiple providers is obtained as
\begin{equation}\label{equ:surplusmsp}
\begin{aligned}
S_{msp}=&\sum_{t=1}^{\tau} {\mathbb{E}\left(U(T\min\{\sum_{i=1}^{I} X_{i}[t],D[t]\})\right)}-\\
&\sum_{i=1}^{I} \delta_{i} \cdot \mu_{95}(X_{i}[t]).
\end{aligned}
\end{equation}


\begin{figure}[t]
\centering
\subfigure[]
{	\label{fig:en}\centering
	\includegraphics[width=0.8\linewidth]{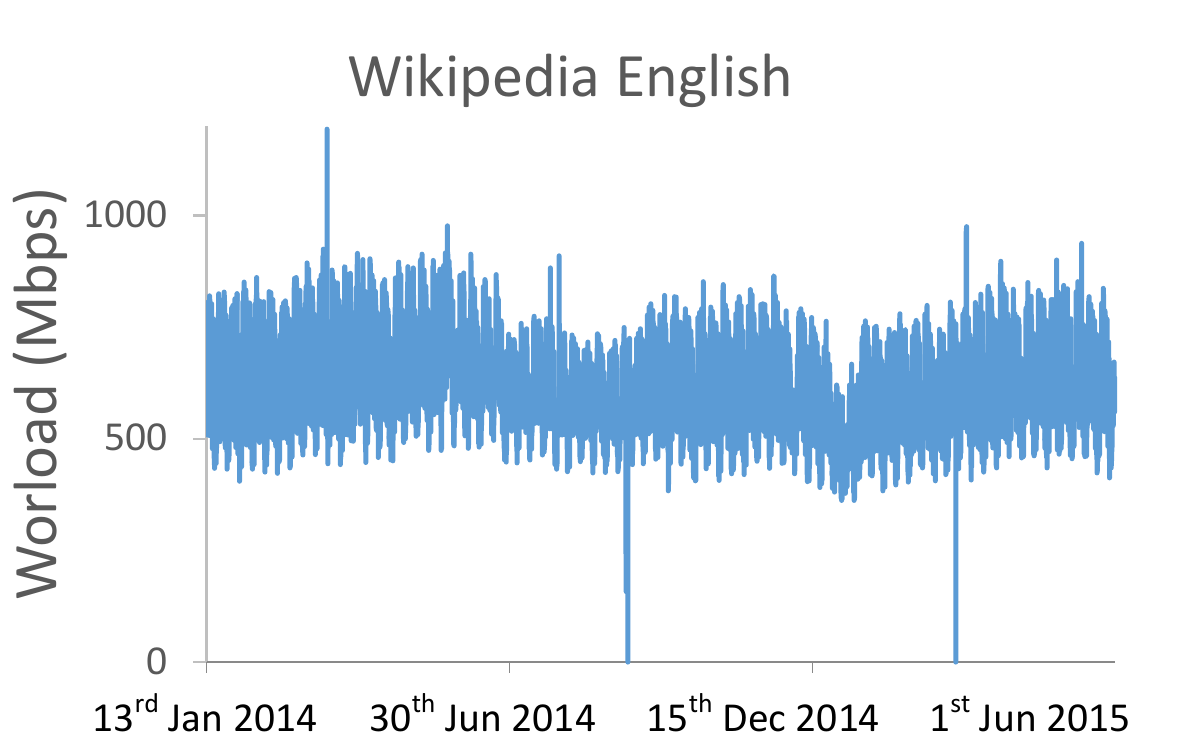}
}
\subfigure[]
{	\label{fig:mw}\centering
	\includegraphics[width=0.8\linewidth]{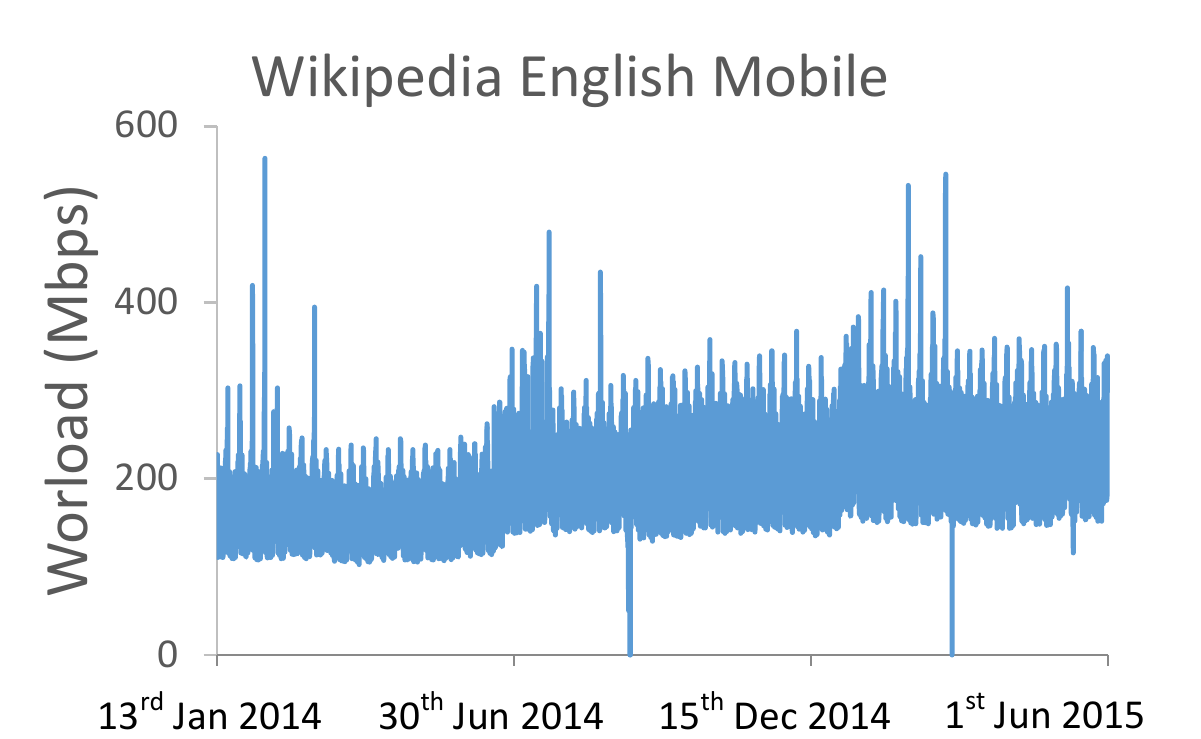}
}
\caption{Examples for the real-world workload traces used in this paper from \cite{Wikipageview}; a) data trace of Wikipedia English, b) data trace of Wikipedia English Mobile.}\label{fig:workload}
\end{figure}
We can also formulate a user's optimization of the usage at  multiple providers to achieve  maximum expected surplus under stochastic prediction of the  demand $ D[t]$ by:
\begin{equation}\label{equ:surplusexpecmsp}
\begin{aligned}
& \underset{X_{i}[t],\rho_{i}[t]}{\text{max}} \!\!\!\!\!\!\
& &  \sum_{t=1}^{\tau} {\sum_{k=1}^{K_{t}} {\pi_{k,t} U(T\min\{\sum_{i=1}^{I} X_{i}[t],D_k[t]\})}}-\\
&&& \sum_{i=1}^{I} \delta_{i}  \max_{t} \rho_{i}[t]X_{i}[t]\\
& \text{s.t.} & & X_{i}[t]\geq 0, && \forall t,i, \\
&&& \rho_{i}[t] \in \{0,1\}, && \forall t,i, \\
&&& \sum_{t=1}^{\tau} {\rho_{i}[t]}=\lceil0.95\tau\rceil, && \forall i. \\
\end{aligned}
\end{equation} 

A special case of problem in (\ref{equ:surplusexpecmsp}) is where  $\forall t$, $K_{t}=1$ and $\pi_{k,t}=1$, i.e.,  the case where the  prediction of user's demand is deterministic and problem (\ref{equ:surplusexpecmsp}) reduces to a deterministic optimization. 
Note that, for the case of multiple providers, the exact solution of the optimization in (18), even for the deterministic optimization, cannot be obtained from the method discussed in Theorem \ref{theorem2}.  
Therefore, for the solution of the problem in (18), as in Section \ref{subsec:stochasticproblem}, we  transform the nonlinear mixed-integer programming (\ref{equ:surplusexpecmsp}) into an equivalent mixed-integer convex programming (as shown in Theorem \ref{theorem3}) or MILP (as shown in Theorem \ref{theorem4}), where the mixed-integer convex programming and the MILP can be solved via CBB and MILP solvers, respectively.

%
%

\subsection{Updating Usage of  Bandwidth  During a  Cycle}\label{subsec:after}

Next, we show that the user can further improve its surplus \emph{during} a billing cycle, by updating its \emph{planned usage} of bandwidth at each time slot based on the newly \emph{exposed demand}. We also show that the user's \emph{final surplus} after a billing cycle  will be no less  than the  expected surplus.
%
Here, we assume that, at the beginning of each time slot, the user's demand for bandwidth  is \emph{exposed} to the user. We denote the \emph{exposed demand} value at time slot $t$ by $\bar{D}[t]$.

Generally, the demand $D[t]$ may not be the same as the exposed value $\bar{D}[t]$. 
Therefore, a user can \emph{update} its \emph{planned usage} of bandwidth in real-time based on the newly learned \emph{exposed demand} information, i.e.,  $\bar{D}[t]$, to further improve its surplus while keeping its bandwidth cost unchanged. For example, if $X[t]< \bar{D}[t]$ and $X[t] <\mu_{95}(X[t])$, the user can increase its usage from $X[t]$ to $\min \{\bar{D}[t],\mu_{95}(X[t])\}$. In this way, the user's net utility can be enhanced while remaining its bandwidth cost unchanged.

In practice, the  \emph{expected} \emph{95th percentile usage} $\mu_{95}(X[t])$ is treated as a rate limiter. According to Theorem \ref{theorem1}, when $\rho[t]=1$, the user restricts its usage at this times slot to reduce its \emph{95th percentile usage}. Specifically, when $\rho[t]=1$, if $\bar{D}[t] \leq \mu_{95}(X[t])$, the user can utilize bandwidth on-demand, and if $\bar{D}[t] > \mu_{95}(X[t])$, the user needs to restrict its  utilization of bandwidth to ensure that its \emph{95th percentile usage} equals to $\mu_{95}(X[t])$. On the contrary, the user can always utilize bandwidth on-demand when $\rho[t]=0$ since the usage at this time slot has no impact on the \emph{95th percentile usage}.
Therefore, we formulate the user's \emph{updated usage} of bandwidth at each time slot during a cycle, which is denoted by $\bar{X}[t]$, as
\begin{equation}\label{equ:realusage}
\bar{X}[t]=
\begin{cases}
\bar{D}[t], &  \text{if} \ \rho[t]=0 \ \text{or} \ \bar{D}[t]\leq \mu_{95}(X[t]);\\
\mu_{95}(X[t]), & \text{otherwise}.
\end{cases}
\end{equation}

From (\ref{equ:realusage}), we ensure that $\forall t$, $\bar{X}[t]\leq \bar{D}[t]$. Similar to (\ref{equ:net utility}) and (\ref{equ:surplus}), after a billing cycle, the  net utility with \emph{updated usage} values $\bar{X}[1],\ldots,\bar{X}[\tau]$ can be calculated as
\begin{equation}
\label{equ:realnet utility}
\bar{R} = \sum_{t=1}^{\tau} {U(T\bar{X}[t])}.
\end{equation}
Further, from  (\ref{mu_95_Model}), (\ref{C_95_Model})  and (\ref{equ:realnet utility}), we formulate the user's  surplus  with \emph{updated usage} values $\bar{X}[1],\ldots,\bar{X}[\tau]$ via
\begin{equation}\label{equ:realsurplus}
\bar{S}=\sum_{t=1}^{\tau} {U(T\bar{X}[t])}-\delta \cdot \mu_{95}(\bar{X}[t]).
\end{equation}

We can show that a user's surplus with \emph{updated usage} values $\bar{X}[1],\ldots,\bar{X}[\tau]$ is always no less than its surplus with \emph{planned usage} values $X[1],\ldots,X[\tau]$.
From  (\ref{equ:realusage}), we ensure that
$\mu_{95}(\bar{X}[t]) \leq \mu_{95}(X[t])$.
Therefore, the bandwidth cost over a billing cycle with  $\bar{X}[t]$ is always no higher than the bandwidth cost with  $X[t]$.
\noindent

Next, we notice that the net utility over a billing cycle with \emph{updated usage} values, $\bar{X}[t]$, is always no less than the bandwidth cost with \emph{planned usage} values, $X[t]$, i.e.,
\begin{equation}\label{proof5.2}
U(T\min\{\bar{X}[t],\bar{D}[t]\}) \geq U(T\min\{X[t],\bar{D}[t]\}), \quad \forall t.
\end{equation}

\noindent To verify that (\ref{proof5.2}) indeed holds,  consider three cases:

\noindent
\textbf{Case 1}: If $\rho[t]=0$, $\bar{X}[t]=\bar{D}[t]$. Since the net utility function $U(\cdot)$ is nondecreasing and $T>0$, (\ref{proof5.2}) is satisfied.
 
\noindent
\textbf{Case 2}: If $\rho[t]=1$ and $\bar{D}[t] \leq \mu_{95}(X[t])$, $\bar{X}[t]=\bar{D}[t]$. Same as case 1, in this case,  (\ref{proof5.2}) is  satisfied.

\noindent
\textbf{Case 3}: If $\rho[t]=1$ and $\bar{D}[t] > \mu_{95}(X[t])$, $\bar{X}[t]=\mu_{95}(X[t])$ and $X[t] \leq \mu_{95}(X[t])$. In this case, (\ref{proof5.2}) is also satisfied.

\noindent

Accordingly, as the surplus of a user over a  cycle equals its net utility minus its bandwidth cost over that cycle, We can see that a user's surplus with \emph{updated usage} values is always no less than its surplus with \emph{planned usage} values. 



\vspace{0.1cm}

Identically, if the user can receive service from multiple providers, we can also update its \emph{planned} usage of bandwidth at provider $i$, i.e., $X_i[t]$, in real-time based on the newly learned information of the  \emph{exposed} demand $\bar{D}[t]$. Let $\bar{X}_{i}[t]$ denote the \emph{updated usage} of bandwidth at provider $i$ at time slot $t$ and it is defined as
\begin{equation} \label{equ:realusagemsp}
\bar{X}_{i}[t]=
\begin{cases}
\bar{D}[t], &  \text{if} \ \rho_{i}[t]=0 \ \text{or} \ \bar{D}[t]\leq \mu_{95}(X_{i}[t]); \\
\mu_{95}(X_{i}[t]), & \text{otherwise},
\end{cases}
\end{equation}
where $\rho_{i}[t]$ is the auxiliary variable as used in Theorem \ref{theorem1}. Then, we formulate the user's   surplus over a  cycle  via
\begin{equation}\label{equ:finalsurplusmsp}
\bar{S}_{msp}=\sum_{t=1}^{\tau} {U(T\sum_{i=1}^{I} \bar{X}_{i}[t])}-\sum_{i=1}^{I} \delta_{i} \cdot \mu_{95}(\bar{X}_{i}[t]).
\end{equation}
Similarly, a user can also further improve its surplus via updating its \emph{planned} usage according to (\ref{equ:realusagemsp}) if it can receive service from multiple providers.

Note that, since the final surplus  a user can achieve in  our design is obtained from (\ref{equ:realsurplus}) and (\ref{equ:finalsurplusmsp}), we use these values as the user's surplus, in the rest of this paper, to evaluate the performance of our design.

\section{Case Studies}
In this section, with real-world data traces, we first study the computation time and performance of our proposed solution methods for solving the stochastic problem (\ref{equ:offsto}). Second, we evaluate our design with a simple method to forecast the demand for bandwidth. Third, we discuss the impact of price and utility factor on the performance of our design. Forth, we show that, with multiple providers, the user can further improve its surplus with our design.
\begin{figure}[!t]
	\centering
	\includegraphics[width=0.82\linewidth]{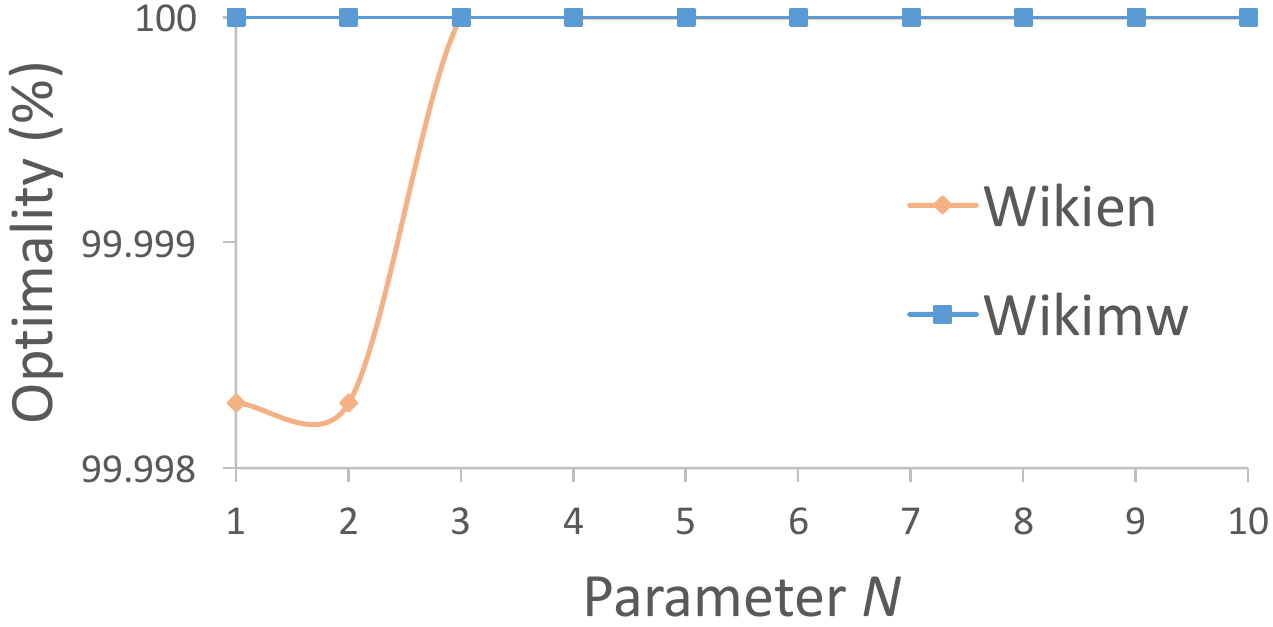}
	\caption{The impact of the number of tangents lines on the optimality of the solution for MILP-based  problem (\ref{equ:offstomilp}).}\label{fig:impactN}
\end{figure}


\subsection{Setup}\label{subsec:evaluationsetup}

We use two data sets in our case studies: 1) \emph{Wikien}: the page view data of Wikipedia English from January 2014 to May 2015 \cite{Wikipageview}, 2) \emph{Wikimw}: the page view data of Wikipedia English Mobile from January 2014 to May 2015 \cite{Wikipageview},  Example traces of these data sets are shown in Fig. 3. Each time slot takes one hour and the billing cycle takes 28 days for Wikien and Wikimw data sets.


The utility functions are selected as follows:
\begin{equation}\label{equ:utilityfunction}
U(x)=
\begin{cases}
A(1-a)^{-1}x^{1-a}, &\mbox{if $a \in(0,1)$};\\
Alog(x), &\mbox{if $a=1$},
\end{cases}
\end{equation}
which is commonly used in economics \cite{Joe-Wong2012, Nicholson2011}. Here, $A>0$ is the utility factor decided by the user and $a\in(0,1]$ measures the concavity of the user's utility. Namely, as $a$ increases, the user's utility becomes more concave. Specifically, we assume that $a=0.1$, $A=0.08$ and the impact of the utility factor $A$ on the surplus of user will be discussed in Section \ref{subsec:impactprice}.


We use a very simple workload forecasting method. Let $D_1[t]$ and $D_2[t]$ denote the workload at time slot $t$ in the last two billing cycles, respectively. Suppose that $\pi_{1,t} = \pi_{2,t} = 0.5$, $\forall t = 1,\ldots,\tau$. Specifically, for deterministic surplus maximization, we assume that $D[t]=\pi_{1,t}D_1[t]+\pi_{2,t}D_2[t]$, for any $t = 1,\ldots,\tau$.

\subsection{Computation Complexity of Proposed Solution Methods}\label{subsec:impacttengentslines}
\begin{figure}[!t]
\centering
\subfigure[]
{	\label{fig:bbruntime}\centering
	\includegraphics[width=0.82\linewidth]{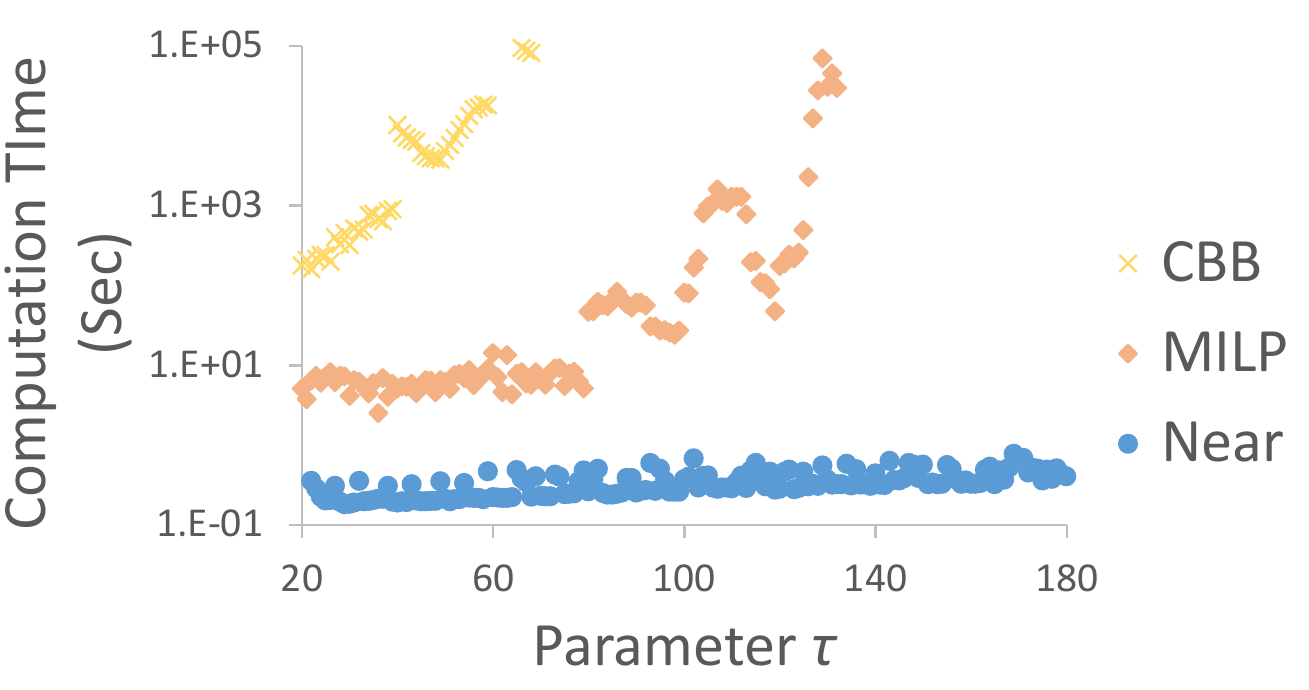}
}
\subfigure[]
{	\label{fig:bbresult}\centering
    \includegraphics[width=0.82\linewidth]{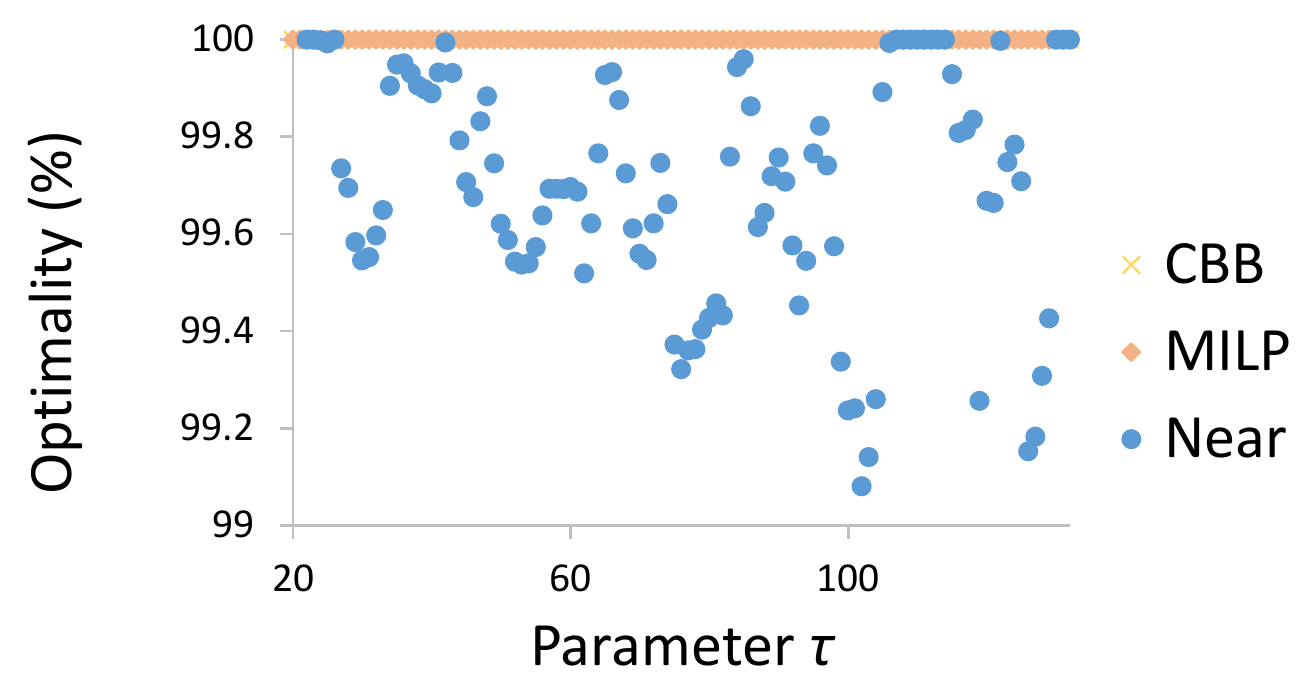}
}
\caption{Comparing different solution methods in solving problem (\ref{equ:offsto}): (a) Computation time, (b) Optimality.}\vspace{-0.35cm}\label{fig:near}
\end{figure}

Recall from Section \ref{subsec:stochasticproblem} that there are multiple options to solve the stochastic problem (\ref{equ:offsto}). Specifically, the proposed CBB method leads to the exact optimal solution. The efficiency of the MILP method, however, depends on the number of tangent lines $N$. Suppose we choose $\Delta[t]=TD_k[t]/N$. Fig. \ref{fig:impactN} shows the optimality in percentage in applying the MILP method versus the number of tangent lines $N$ for different datasets. We can see that the results are accurate when $N\geq 3$. Therefore, for the rest of this paper, we assume that $N=3$.





Next, we evaluate the computation time for each solution method. We use a personal computer with Intel Xeon CPU E5-2450 @2.50GHZ. The results are shown in Fig. \ref{fig:bbruntime}. We can see that the computation time of CBB is much longer than MILP. Even for the MILP approach, it may take several hours to find the global optimal solution of problem (\ref{equ:offstomilp}) as the size of the problem increases.

As we pointed out in Section \ref{subsec:stochasticproblem}, one can obtain an \emph{approximate} solution for problem (\ref{equ:offstomilp}) by terminating the optimization solver at certain guaranteed optimality bounds. This can be done by setting up a stopping condition for the MILP method based on the ratio between the upper-bound and the lower-bound solutions. The upper-bound solution is the surplus that can be achieved if we relax the remaining binary variables at the current branching stage. The lower-bound solution is the surplus at the best binary solution that has been obtained so far at the current branching stage. Clearly, this ratio indicates a guaranteed optimality in the  solution of MILP that has already been reached at the current branching stage. In this paper, we obtain an approximate solution by stopping the MILP method in CPLEX once the above mentioned ratio reaches 5\%, which guarantees at least 95\% optimality. We refer to this approximate solution approach as the \emph{Near} method.

As we can see in Fig. \ref{fig:bbruntime}, the Near method is significantly less  complex in terms of required computation,  compared to the CBB and MILP methods. Specifically, the computational time for the Near method grows only linearly with respect to the number of time slots. Interestingly, we can see in Fig. \ref{fig:bbresult} that the actual achieved optimality is around 99\% or more, i.e., much better than the guaranteed 95\% worst case optimality value.
Therefore, for the rest of this paper, we use the Near method at $95\%$ guaranteed optimality.



\subsection{Performance Evaluation}\label{subsec:performanceevaluation}

\begin{figure}[!t]
\centering
\subfigure[]
{	\label{fig:enperformance}\centering
	\includegraphics[width=0.8\linewidth]{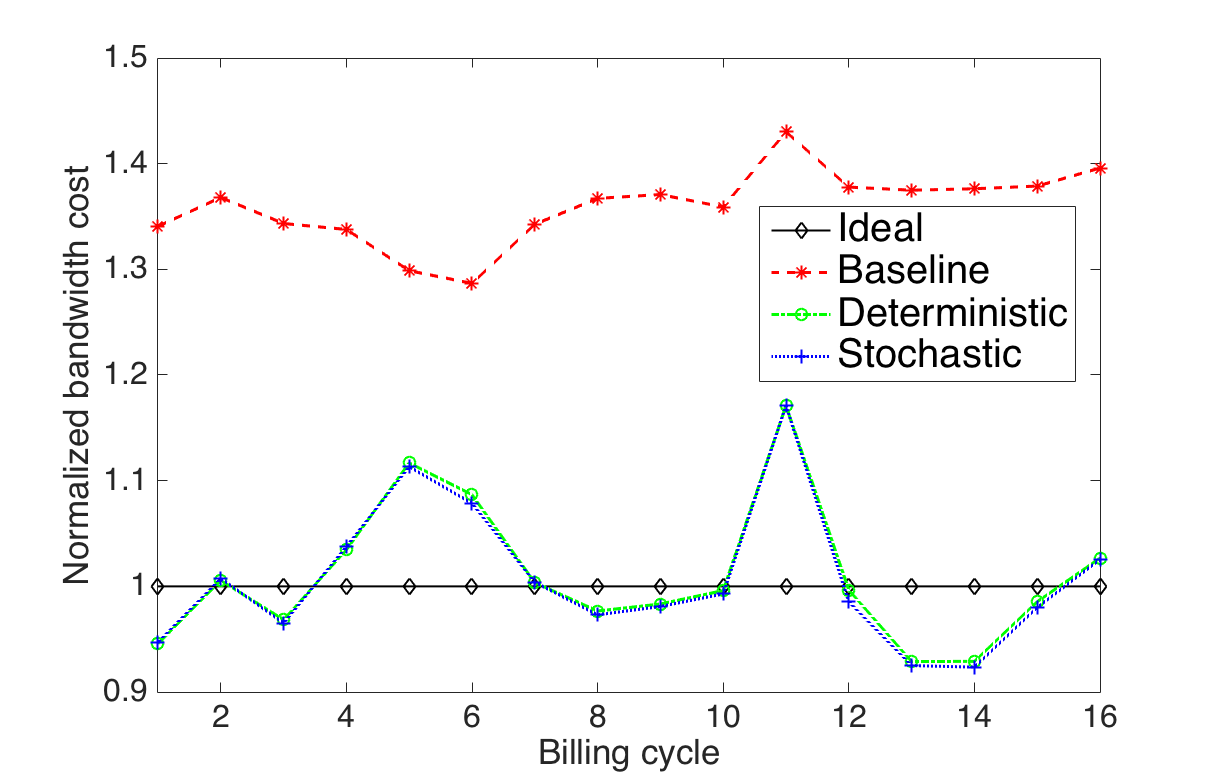}
}
\subfigure[]
{	\label{fig:mwperformance}\centering
	\includegraphics[width=0.8\linewidth]{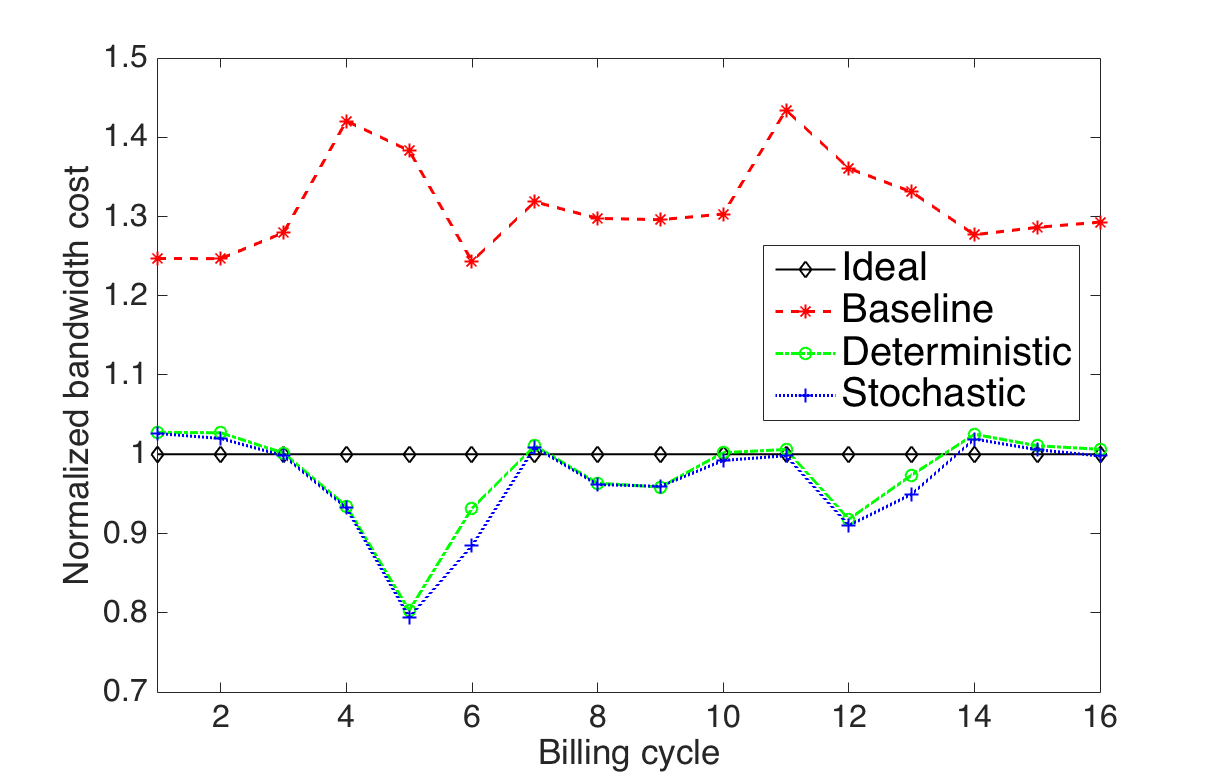}
}
\caption{Comparing normalized bandwidth cost under different methods and different workloads: a) Wikien, b) Wikimw.\vspace{-0.2cm}}\label{fig:bandwidthcost}
\end{figure}

\begin{figure}[!t]
\centering
\subfigure[]
{	\label{fig:wikienperformance}\centering
	\includegraphics[width=0.8\linewidth]{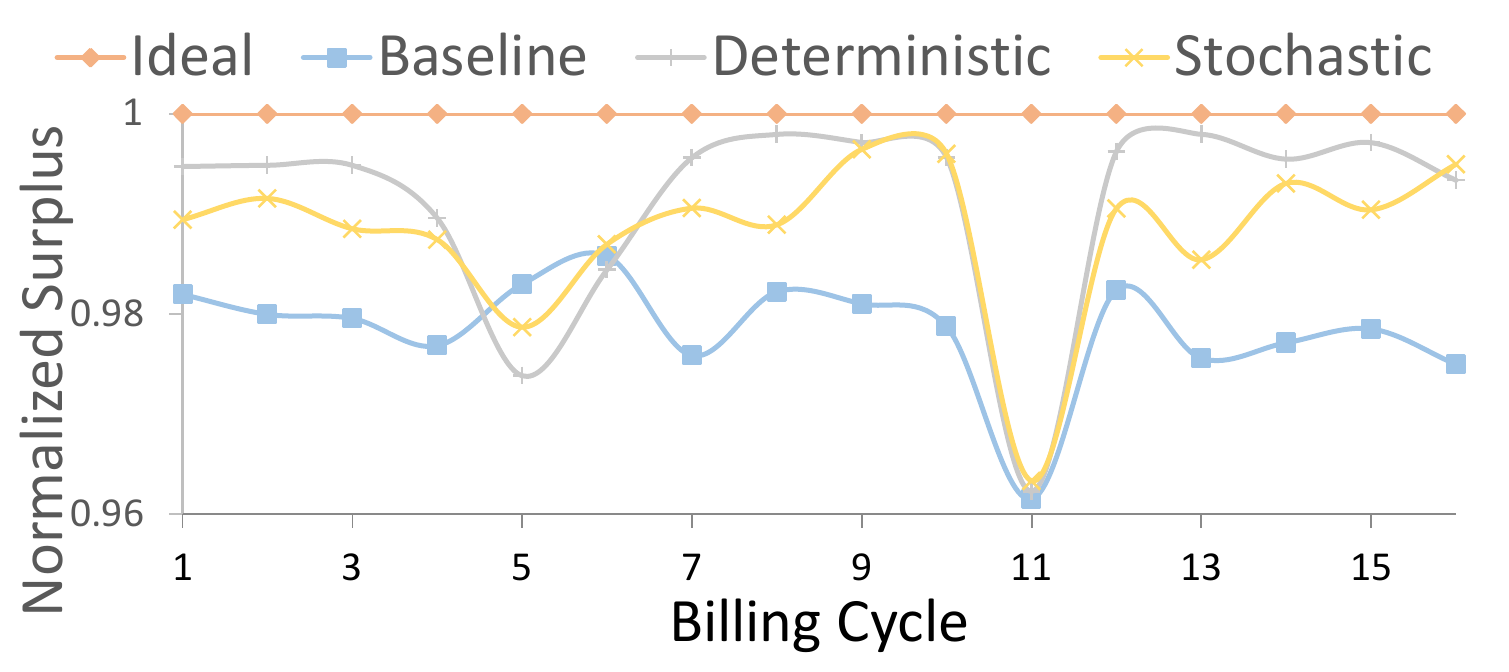}
}
\subfigure[]
{	\label{fig:wikimwperformance}\centering
	\includegraphics[width=0.8\linewidth]{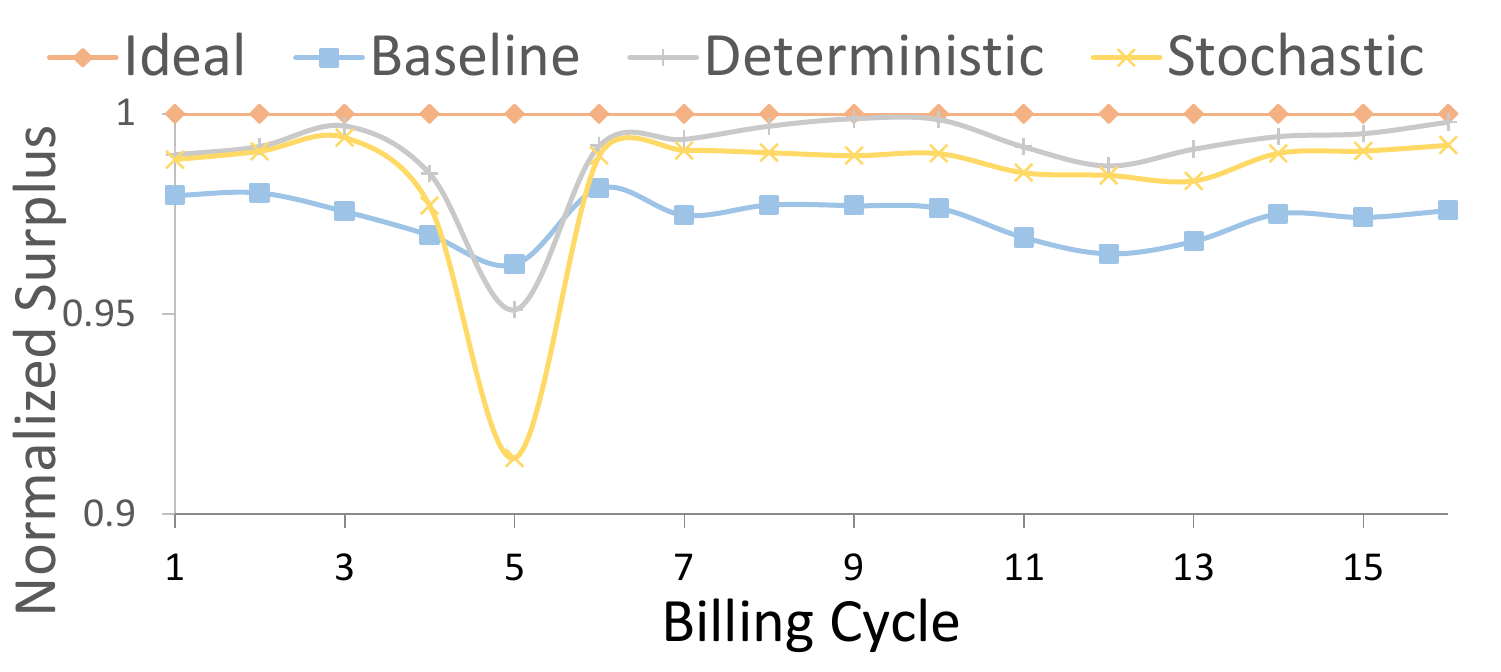}
}
\caption{Comparing normalized surplus under different methods and different workloads: a) Wikien, b) Wikimw.\vspace{-0.2cm}}\label{fig:performance}
\end{figure}

\begin{figure}[!t]
\centering
\subfigure[]
{	\label{fig:wikienaverage}\centering
	\includegraphics[width=0.8\linewidth]{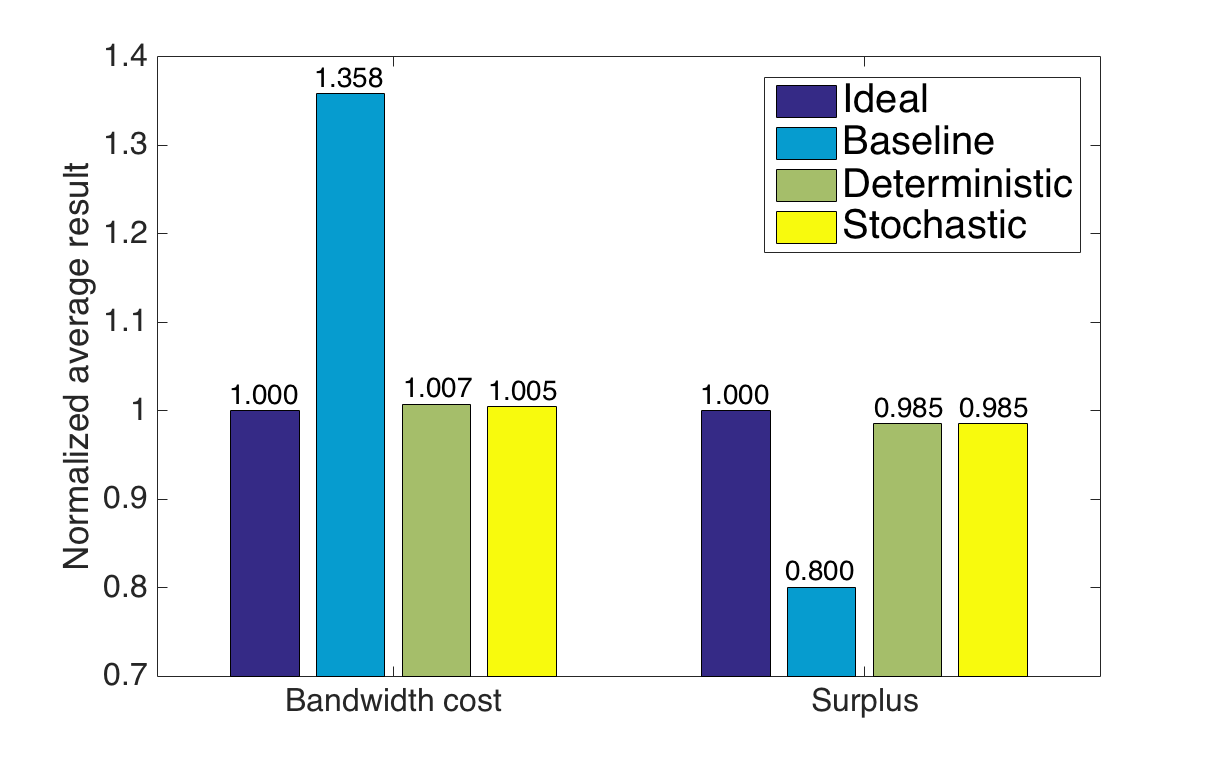}
}
\subfigure[]
{	\label{fig:wikimwaverage}\centering
	\includegraphics[width=0.8\linewidth]{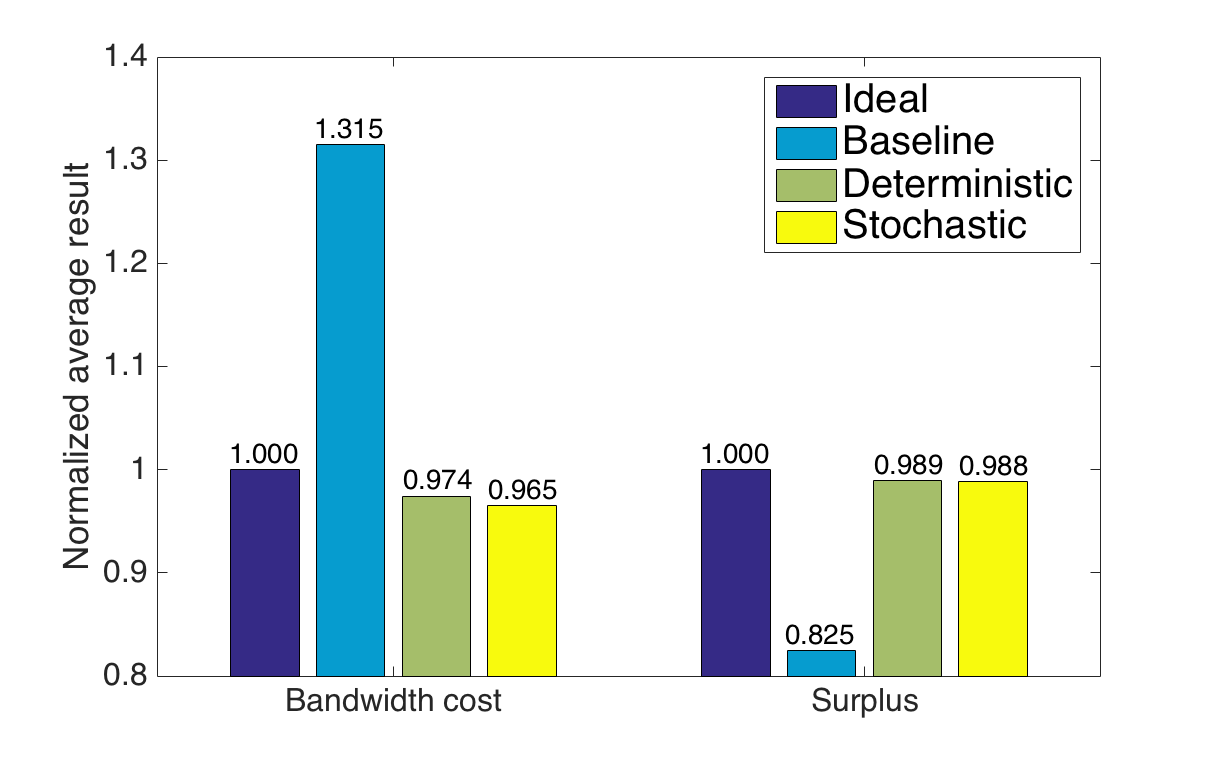}
}
\caption{Comparing average bandwidth cost and surplus under different methods and different workloads: a) Wikien, b) Wikimw.\vspace{-0.2cm}}\label{fig:average}
\end{figure}
As a \emph{Baseline} for performance comparison, we consider the case where the bandwidth is allocated on-demand, i.e., $X[t]=\bar{X}[t]=\bar{D}[t]$, for any $t = 1, \ldots, \tau$. Note that, this approach resembles how the bandwidth is currently allocated in practice. Next, we also assume an \emph{Ideal} case where the usage of bandwidth is optimized based on \emph{true knowledge} of demand, i.e., $\forall t$, $D[t]=\bar{D}[t]$. While the Baseline shows how well we can perform compared to the existing practice, the Ideal case shows the best performance that we can ever get, assuming that we can perfectly predict the upcoming workload.


Next, we compare the Baseline and Ideal cases with our proposed \emph{Deterministic} and \emph{Stochastic} methods. The Deterministic method refers to the case where the bandwidth usage is scheduled based on the optimal solution of the deterministic surplus maximization problem in (\ref{equ:deter}). The Stochastic method refers to the case where the bandwidth usage is scheduled based on the optimal solution of the stochastic surplus maximization problem in (\ref{equ:offstomilp}) using the Near method with $95\%$ guaranteed optimality. The method of forecasting the workload in each case was already explained in Section \ref{subsec:evaluationsetup}.

The results on performance comparison are shown in Fig. \ref{fig:bandwidthcost}, Fig. \ref{fig:performance} and Fig. \ref{fig:average}, where the results for all methods are normalized with respect to the results of the Ideal case. 
Here, the price of bandwidth is set to be  \$15 per Mbps. We can make the following observations based on these results:
%
%
%
\begin{itemize}

  \item As shown in Fig. \ref{fig:bandwidthcost} and Fig. \ref{fig:performance}, even though we use a very simple method to forecast the demand for bandwidth, the Deterministic and Stochastic solutions outperform the Baseline in both bandwidth cost reduction and surplus improvement. Thus, our method is robust to the error of prediction of user's demand. Meanwhile, Deterministic and Stochastic have similar outcomes. 
      \vspace{0.05cm}

 \item As shown in Fig. \ref{fig:average}, on average, our proposed optimization-based approach to respond to burstable billing can greatly reduce the user's bandwidth cost while improving its surplus when comparing against Baseline. For example,with data trace of Wikien, both Deterministic and Stochastic surplus maximization can reduce the user's bandwidth cost by $26\%$ while increasing its total surplus by $23\%$, respectively.
\end{itemize}

      \vspace{0.05cm}
\begin{figure}[!t]
\centering
\subfigure[]
{	\label{fig:ImpactPriceWikien}\centering
	\includegraphics[width=0.8\linewidth]{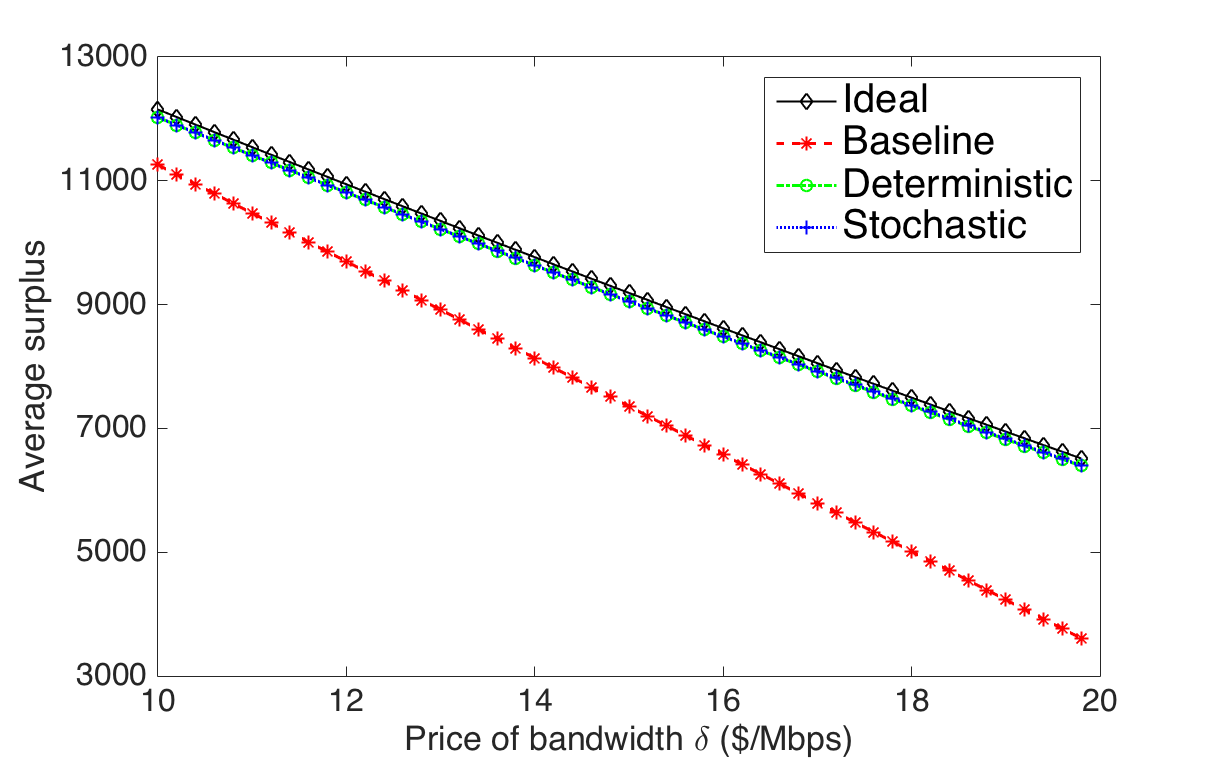}
}
\subfigure[]
{	\label{fig:ImpactPriceWikimw}\centering
	\includegraphics[width=0.8\linewidth]{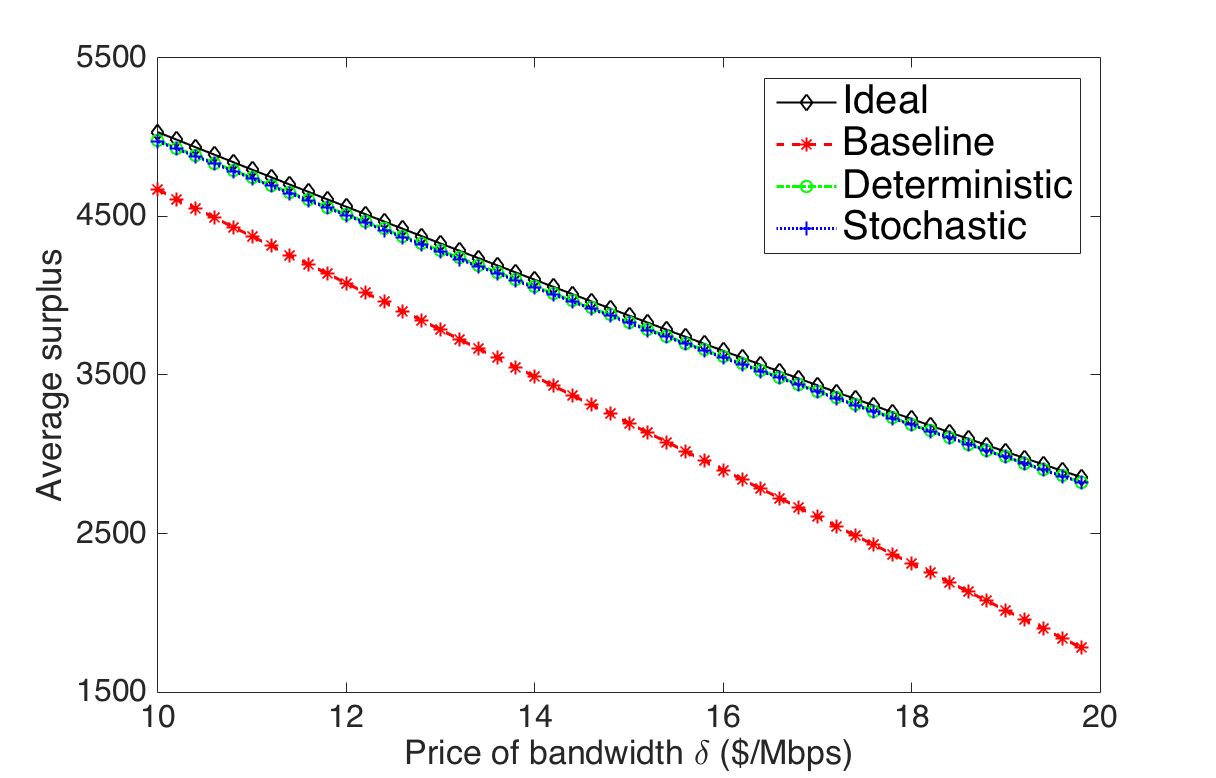}
}
\caption{The impact of the price of bandwidth on average surplus under different workloads: a) Wikien, b) Wikimw.}\label{fig:impactprice}
\end{figure}
\begin{figure}[!t]
\centering
\subfigure[]
{	\label{fig:ImpactAWikien}\centering
	\includegraphics[width=0.8\linewidth]{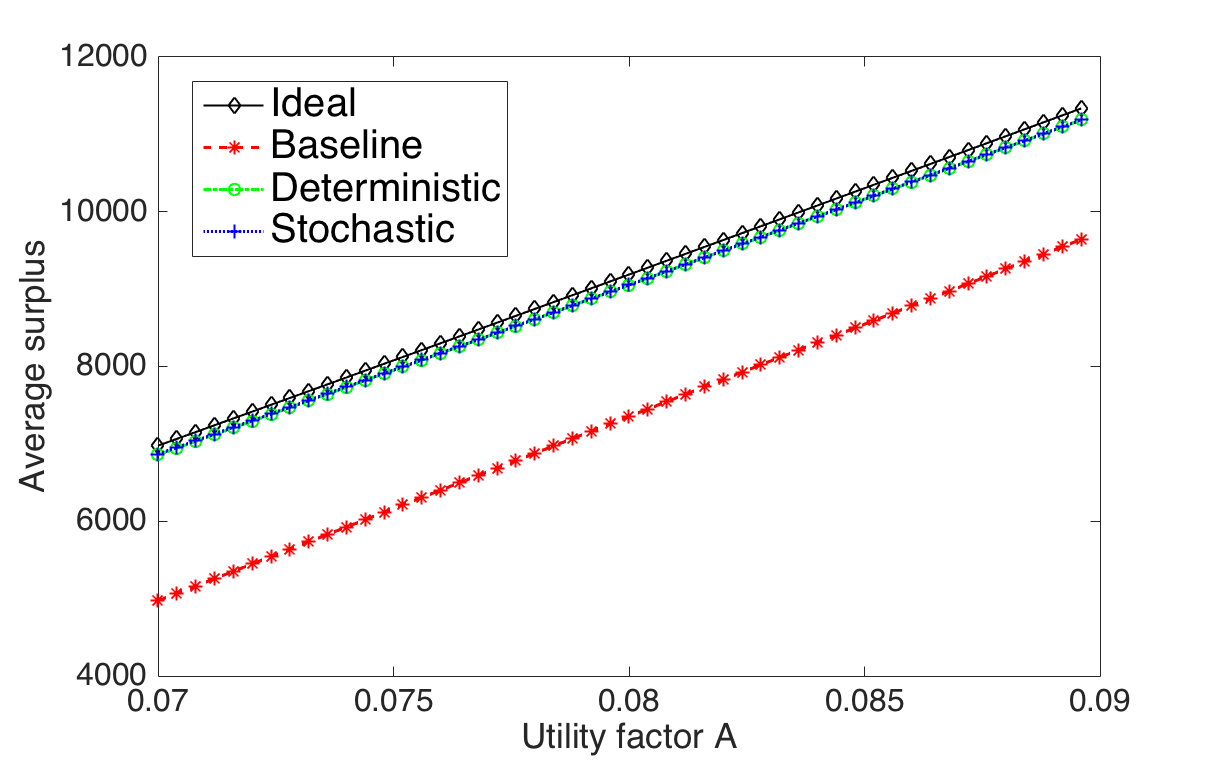}
}
\subfigure[]
{	\label{fig:ImpactAWikimw}\centering
	\includegraphics[width=0.8\linewidth]{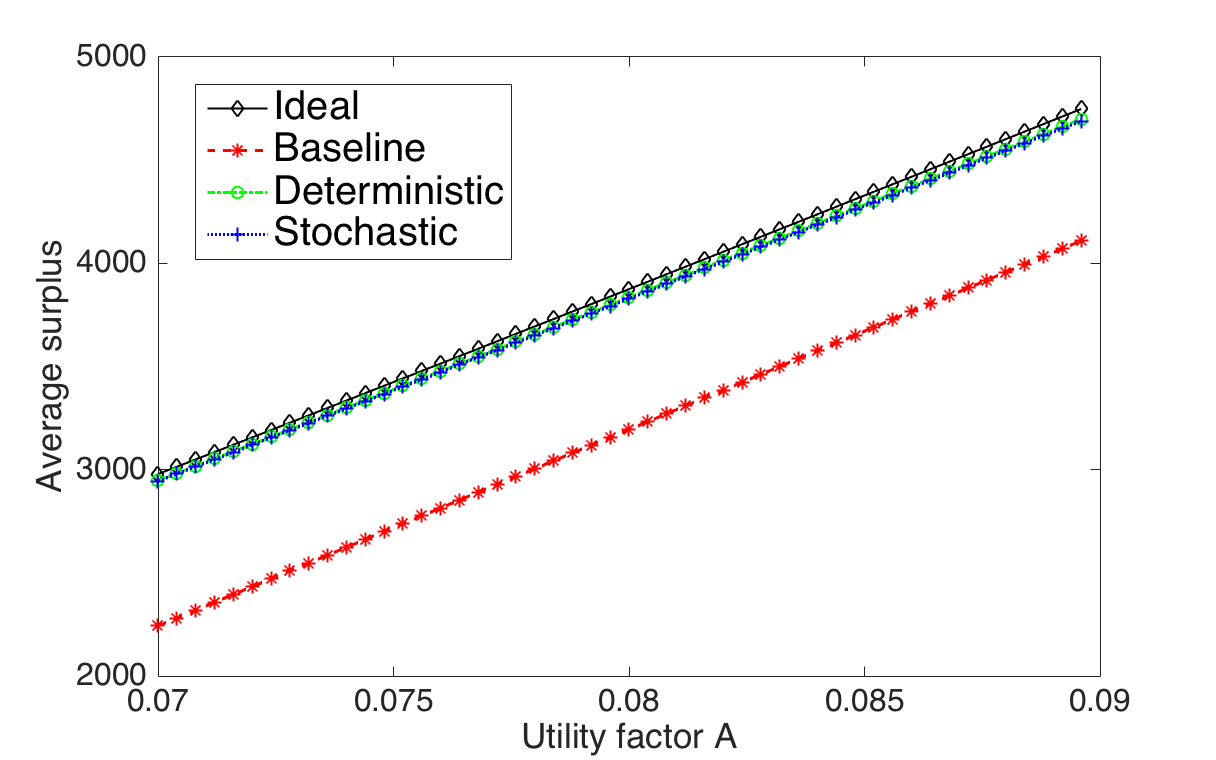}
}
\caption{The impact of the utility factor on average surplus under different workloads: a) Wikien, b) Wikimw.}\label{fig:impactA}
\end{figure}

\begin{figure}[!t]
\centering
\subfigure[]
{	\label{fig:mulwikienbandwidth}\centering
	\includegraphics[width=0.8\linewidth]{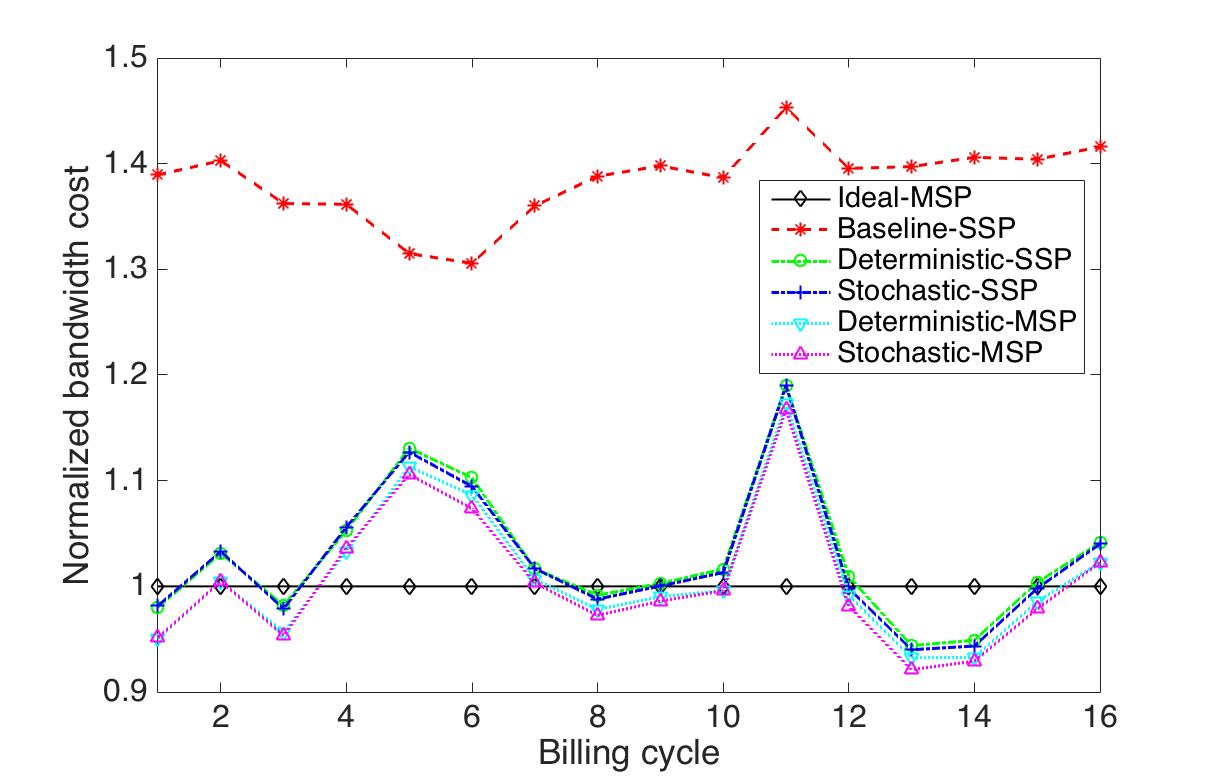}
}
\subfigure[]
{	\label{fig:mulwikimwbandwidth}\centering
	\includegraphics[width=0.8\linewidth]{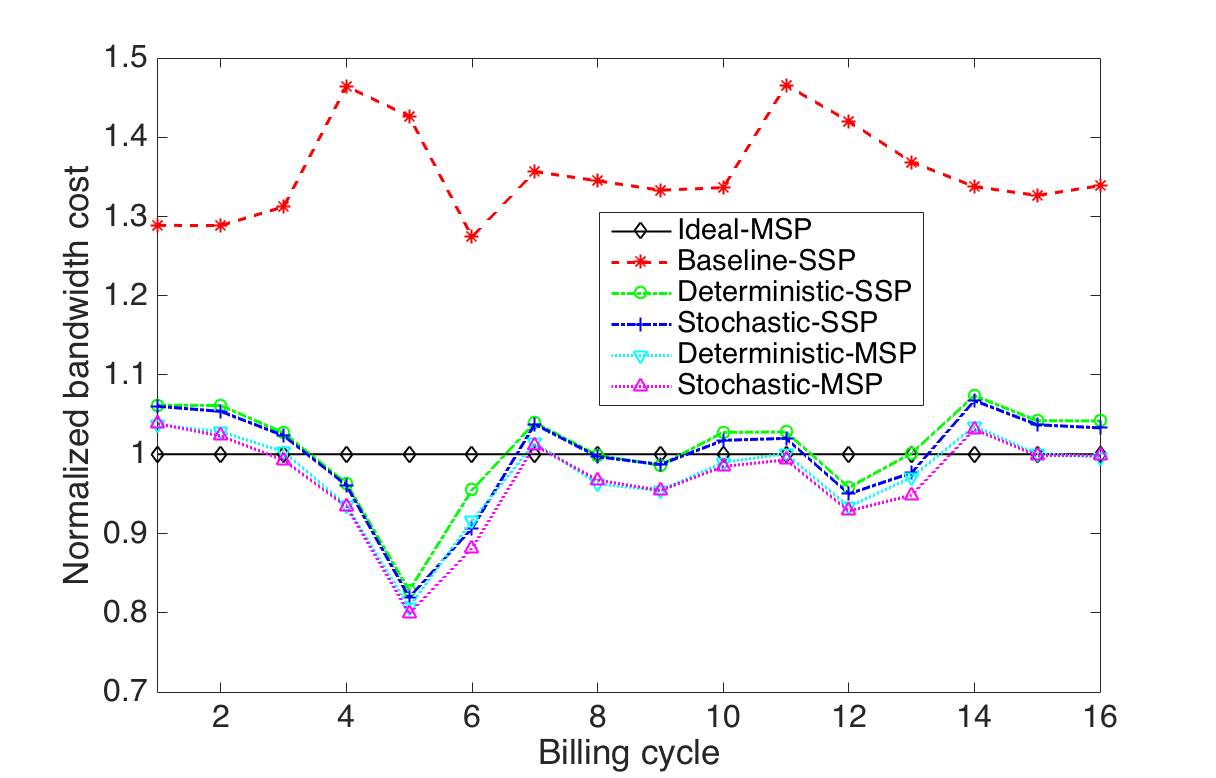}
}
\caption{Comparing normalized bandwidth cost with multiple providers under different workloads: a) Wikien, b) Wikimw.}\label{fig:bandwidthmul}
\end{figure}

\begin{figure}[!t]
\centering
\subfigure[]
{	\label{fig:mulwikienperformance}\centering
	\includegraphics[width=0.8\linewidth]{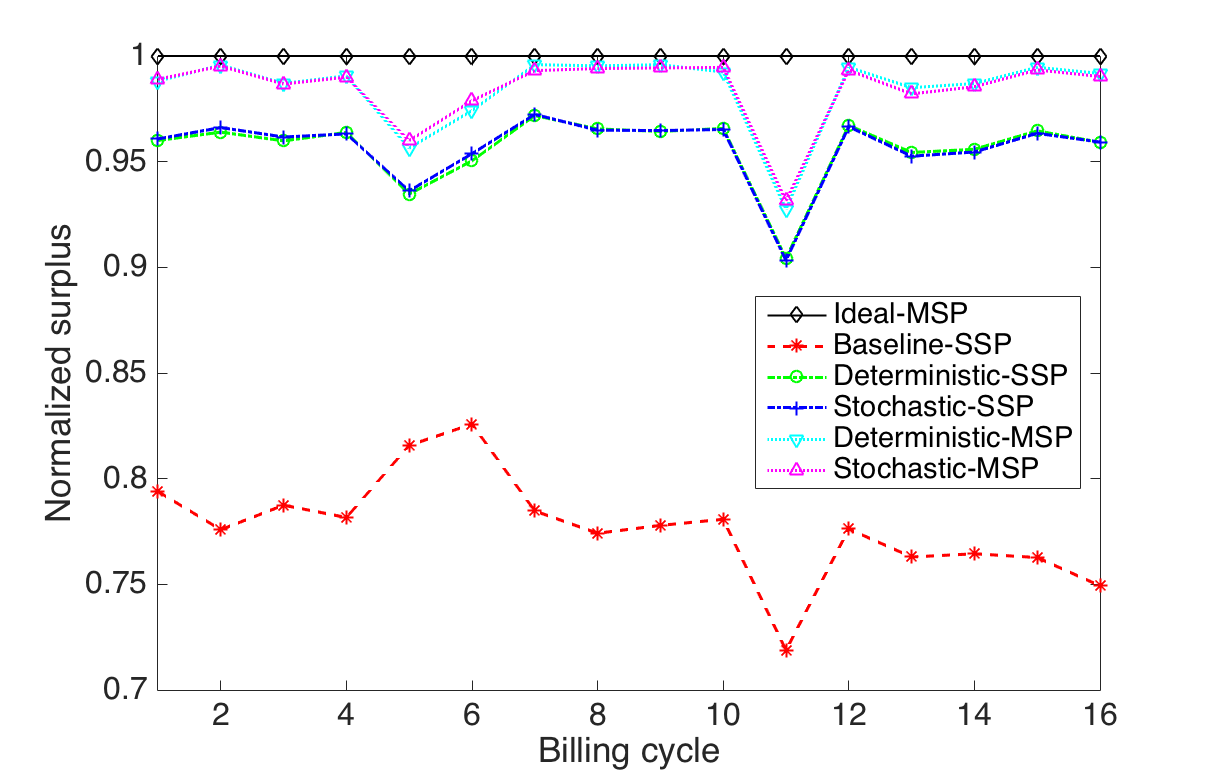}
}
\subfigure[]
{	\label{fig:mulwikimwperformance}\centering
	\includegraphics[width=0.8\linewidth]{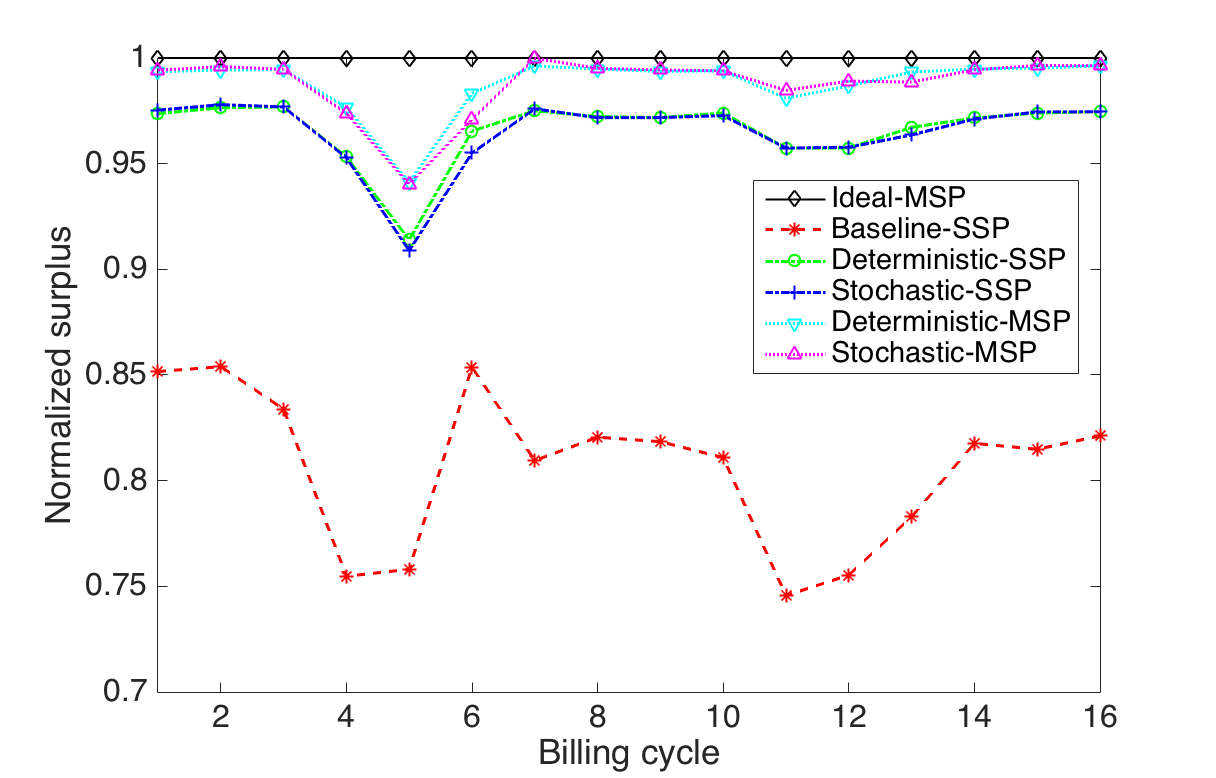}
}
\caption{Comparing normalized surplus with multiple providers under different workloads: a) Wikien, b) Wikimw.}\label{fig:performancemul}
\end{figure}
\subsection{Impact of Price and Utility Factor}\label{subsec:impactprice}
Intuitively, increasing the price for bandwidth would increase the user's cost. Accordingly, the surplus that the user may gain decreases as we increase price parameter $\delta$. However, the rate of such decrease is \emph{not} the same for different methods. The results are shown in Fig. \ref{fig:impactprice}. We can see that the rate of decrease in surplus is higher for the Baseline compared to the Deterministic and Stochastic methods. As a results, the surplus improvements with our proposed optimization-based approaches are higher when the price of bandwidth is high. 

Next, we analyze the impact of utility factor $A$. Clearly, increasing $A$ results in higher surplus for the same usage of bandwidth. By analysing Fig. \ref{fig:impactA}, we find that the distance between Baseline and Deterministic/Stochastic is slightly larger when $A$ is small. Namely, users with smaller utility factors, who are more sensitive to price than performance, are more likely to response to the burstable billing to improve their surpluses. We can also see that the Deterministic and Stochastic methods outperform the Baseline at all choices of $A$.



\subsection{Impact of Multiple Providers}\label{subsec:impactmulti}
Suppose the user can receive service from two providers, who are referred to as providers $1$ and $2$. Both of them offer bandwidth at \$15 per Mbps. 

To evaluate our proposed approach to response to burstable billing with multiple providers, we simulate six different cases:
\begin{itemize}
\item \emph{Ideal-MSP}: It is defined as the outcome of maximizing surplus, under the assumption that the demand for bandwidth is known with multiple providers.  

\item \emph{Baseline-SSP}: In this case, the user utilizes bandwidth from provider 1 on-demand.

\item \emph{Deterministic-SSP}: In this case, the user utilizes bandwidth from provider 1 and makes its decisions
based on our design with deterministic prediction about its demand. 

\item \emph{Stochastic-SSP}: In this case, the user utilizes bandwidth from provider 1 and makes its decisions
based on our design with stochastic prediction about its demand. 

\item \emph{Deterministic-MSP}: In this case, the user utilizes bandwidth from both provider 1 and 2 and makes its decisions
based on our design with deterministic prediction about its demand. 

\item \emph{Stochastic-MSP}: In this case, the user utilizes bandwidth from both provider 1 and 2 and makes its decisions
based on our design with stochastic prediction about its demand. 
\end{itemize}


Figures  \ref{fig:bandwidthmul} and  \ref{fig:performancemul} show the normalized bandwidth cost and surplus, obtained in six different cases, where the base for normalization is the surplus under the Ideal-MSP case. We can see that Deterministic-MSP and Stochastic-MSP methods always outperform Baseline-SSP in both bandwidth cost reduction and surplus improvement. 
Finally, we  also find that Deterministic-MSP and Stochastic-MSP are always better than Deterministic-SSP and Stochastic-SSP. We  may infer  that the availability of multiple providers further reduce the user's bandwidth cost and improves its surplus under optimal response mechanism to burstable billing.

\section{Conclusion and Future Work}\label{sec:conclusion}
A novel optimization-based approach was proposed to select the usage of bandwidth for a user, such as a user of a colocation data center, who is charged for bandwidth usage under burstable billing. Our proposed approach considers workload demand uncertainty, and is general in the sense that it does not make any assumption about the statistical characteristics of workload.
%
%
Numerical results based on empirical case studies confirm that even with a simply workload forecasting method, the user can obtain significantly higher surplus under the proposed optimal method for responding to burstable billing, compared to the current practice of allocating bandwidth on-demand. We also extended our design to another emerging practical scenario where a user can receive service from multiple providers. Accordingly, besides bandwidth allocation, our problem formulation also addresses workload distribution.

This paper can be extended in several directions. First, one can adopt a more advanced workload forecasting method to better model probability distribution functions for the demand for bandwidth. In fact, with enough accurate prediction, the performance of the proposed methods are guaranteed to improve. Second, one can try to further reduce a user's \emph{95th percentile usage} via traffic shaping \cite{Marcon2011}, traffic aggregation \cite{Stanojevic2011}, traffic shifting in time and space \cite{Clegg2014}, simultaneously. Finally, one can revisit the problem from the provider's viewpoint based on the knowledge of how a user optimally responds to burstable billing and adjusts the billing parameters to achieve better results for the provider.

\appendices
\end{document}